%% file: usenix.tex
\documentclass[letterpaper,twocolumn,10pt]{article}
\usepackage{usenix,epsfig,endnotes}

\usepackage{enumitem}
\usepackage[caption=false]{subfig}
\usepackage{csquotes}

\usepackage{amsmath}
\usepackage{amssymb}
\usepackage{dsfont}

\usepackage{graphicx}
\usepackage{makecell}
\usepackage{multirow}
\usepackage{hyperref}
\usepackage{xcolor}
\usepackage{booktabs,nicematrix}
\usepackage{cleveref}
\usepackage{xspace}
\usepackage{algorithm}
\usepackage{algorithmicx}
\usepackage{algpseudocode}
\usepackage{balance}
\usepackage{placeins}
\usepackage{wrapfig}
\usepackage{fancyhdr}
\usepackage{datetime}
\usepackage[available]{usenixbadges}
\usepackage{newtxtext,newtxmath}

\newcommand{\TR}{Tree-Ring\xspace}

\newcommand{\para}[1]{\vskip 4pt\noindent\textbf{#1}\hskip .05in}

\begin{document}

%don't want date printed
\date{}

%make title bold and 14 pt font (Latex default is non-bold, 16 pt)
\title{\Large \bf A Crack in the Bark: Leveraging Public Knowledge \\ to Remove Tree-Ring Watermarks\thanks{This work has been accepted for publication in the 34th USENIX Security Symposium (USENIX'25).}}

\author{
 {\rm Junhua Lin}
 \qquad
 {\rm Marc Juarez}\\University of Edinburgh
}

\maketitle

% Use the following at camera-ready time to suppress page numbers.
% Comment it out when you first submit the paper for review.
% \thispagestyle{empty}

\subsection*{Abstract}
%\jun{Need to shorten this down, its max 200 words abstract}
%\mjm{I'll take care of it}
We present a novel attack specifically designed against \TR, a watermarking technique for diffusion models known for its high imperceptibility and robustness against removal attacks.
%In this study, we present a novel attack specifically designed against it that removes its watermarks while preserving image quality.
Unlike previous removal attacks, which rely on strong assumptions about attacker capabilities, our attack only requires access to the variational autoencoder that was used to train the target diffusion model, a component that is often publicly available.
By leveraging this variational autoencoder, the attacker can approximate the model's intermediate latent space, enabling more effective surrogate-based attacks. 
Our evaluation shows that this approach leads to a dramatic reduction in the AUC of \TR detector's ROC and PR curves, decreasing from 0.993 to 0.153 and from 0.994 to 0.385, respectively, while maintaining high image quality. Notably, our attacks outperform existing methods that assume full access to the diffusion model.
These findings highlight the risk of reusing public autoencoders to train diffusion models---a threat not considered by current industry practices.
Furthermore, the results suggest that the \TR detector's precision, a metric that has been overlooked by previous evaluations, falls short of the requirements for real-world deployment.

\section{Introduction}
Recent advances in generative AI have led to the development of text-to-image models capable of creating highly realistic content. Some of these, such as Midjourney and DALL-E, are proprietary models that produce images nearly indistinguishable from those generated by humans~\cite{lu2023seeing, kamali2024distinguish}. As these models become more sophisticated, discerning between AI-generated and authentic content becomes increasingly difficult, raising serious concerns about the risks of misuse, such as the potential to create \emph{deepfakes} and spread misinformation~\cite{bird2023typology}.
%\jun{Furthermore, the rise in popularity of text-to-image models have surfaced communities with the explicit purpose of sharing these models (e.g. Hugging Face~\cite{huggingface}, CivitAI~\cite{civitai}) as well as attempts to monetize them as services (e.g. Flux~\cite{flux}, ShuttleAI~\cite{shuttleai}).}

In response, the research community has turned its attention to developing watermarking techniques that the owners of the models can deploy to ensure that AI-generated content can be reliably identified. If these watermarks are robust against removal, they can prevent malicious actors from passing off deepfakes as authentic, thereby helping to mitigate some of the risks associated with the misuse of these technologies.
Recent examples of watermarking schemes developed for this purpose are Stegastamp~\cite{tancik2020stegastamp}, Stable Signature\cite{fernandez2023stable}, and \TR~\cite{wen2023treering}, with \TR standing out for the imperceptibility of its watermarks in the generated images and its robustness against removal attacks.

\TR was specifically designed for diffusion models. It achieves high imperceptibility by encoding a fixed string into the diffusion model's initial latent variable, rather than into the image directly. Embedding a fixed watermark constrains the latent space to a region, altering the model's output distribution; however, individual images generated from the watermarked region are clean images, free of unusual patterns or artifacts. To verify a watermarked image, \TR must accurately approximate the inversion of the diffusion process to recover the latent variable that generated the image and confirm the presence of the watermark.

Evaluations of \TR have shown that attacks based on simple image manipulations, such as rotation and cropping, fail to remove the watermark without substantially modifying the image~\cite{wen2023treering}. Consequently, the attacked images cannot serve as substitutes for the watermarked ones, losing their value from an attacker's perspective.
However, these initial evaluations did not explore attacks that expressly target the latent space, where the watermarks are embedded.
%these manipulations are applied directly in the pixel space, rather than targeting the latent space, where the watermarks are embedded.

%whereas the watermarks are embedded in the latent space, that there is potential
%leaving room for attacks that target the latent representations.
%attacks that rely on basic pixel-level operations overlook the fact that the watermarks are embedded in the latent space, underestimating the potential success of a motivated adversary.

%Building upon this observation,
Subsequent evaluations include more advanced attacks tailored to \TR's latent-space embedding mechanism~\cite{zhao2023invisible,lukas2024leveraging,an2024waves}. These attacks are formulated as a learning problem aimed at shifting the latent away from the watermarked region while bounding image quality degradation.
Although some of these attacks are more effective than the simple attacks evaluated by Wen et al., they have been criticized for making strong assumptions about the adversary's knowledge of the target diffusion model~\cite{an2024waves}. Some attacks even require white-box access to the diffusion model, which would eliminate the need for an attack, as the adversary could simply use the model to generate their own non-watermarked images.

We demonstrate how an attacker can use publicly available auxiliary models to overcome some of the assumptions made in prior work.
In particular, we present an attack inspired by An et al.'s surrogate-detector approach\cite{an2024waves}.
Following their attack strategy, we train a model that mimics \TR's detector and find adversarial examples on it that transfer to the original detector.
However, rather than training the detector on the pixel representation of images, we train it on intermediate latents approximated via a publicly available variational autoencoder (VAE).
Since VAEs are commonly used to reduce the training cost of diffusion models, they are often shared and reused, making them readily available to attackers.
%Diffusion models are often trained with a VAE to reduce computational costs, 
%The trend of reusing and sharing VAEs to reduce the computational cost of training diffusion models makes this approach practically feasible.
%Since diffusion models are often trained with a VAE to reduce computational costs, model developers tend to share and reuse pretrained VAEs---a practice that increases the practical feasibility of our attacks.

Our VAE-enhanced surrogate detector attacks outperform previous attacks, including those of An et al., drastically decreasing \TR's detection performance while maintaining image quality.
This improvement in attack performance highlights the risks associated with sharing and reusing VAEs for training diffusion models, a practice that is currently overlooked in the industry (for example, see \cite{flux,dalle-vae,shuttleai,clipdrop,leonardoai,dreamstudio,aiseoart,davinci_sdxl,cogview4}).

\smallskip
\noindent More specifically, our primary contributions are:

\para{An in-depth analysis of \TR's embedding mechanism.} We identify critical flaws in \TR's embedding mechanism, motivating the design of our attacks. The embedding mechanism violates the assumption that the latents are drawn from a Gaussian distribution. This, coupled with the persistence of the watermarks through the diffusion process exposes discriminating features that enable the training of a surrogate detector in the latent space.

\para{Novel VAE-based surrogate attacks.}
We leverage public VAEs to attack \TR watermarking.
Assuming access to a VAE is realistic because, in practice, a pretrained autoencoder is often used to train diffusion models~\cite{Rombach_2022_stable}, but
we also challenge this assumption by evaluating the attacks using a different VAE.
Our attacks successfully remove \TR watermarks, decreasing the detector's TPR@1\%FPR from 0.968 to 0.04 and its ROC-AUC from 0.993 to 0.15, while retaining high image quality. This is nearly a 5-fold decrease in ROC-AUC over the state-of-the-art surrogate detector attack.

\para{An evaluation of detector precision.}
In contrast to previous evaluations of watermarking schemes, we measure \TR's detection precision under base rates that capture a range of deployment scenarios. In a balanced setting, our attack reduces the PR-AUC from 0.994 to 0.385. The impact is even more pronounced at lower, more realistic base rates, suggesting that \TR's precision may be insufficient for the intended use cases it aims to support.

\section{Background and Related Work}
This section provides the necessary background on diffusion and \TR watermarking to follow the rest of the paper.

\subsection{Diffusion Models}
A diffusion model~\cite{sohl2015deep} is based on a \emph{forward diffusion process}, which involves gradually adding noise to a data point until the signal is destroyed. By learning how to reverse this process, a diffusion model estimates the underlying data distribution and provides a method for sampling new data points from it.

More formally, let $x_0$ be a data point drawn from the true data distribution $q(x_0)$. The forward diffusion process is a Markov chain defined by a sequence of $T$ intermediate latent variables $x_1,...,x_T$ in the same domain as $x_0$, where each $x_t$ is obtained by adding noise drawn from the noise distribution $q(x_t\mid x_{t-1})$ to the previous latent, $x_{t-1}$.

The diffusion model is  a distribution $p_\theta(x_0)$, parametrized by $\theta$, that reverses the forward process thus providing a method to sample from $q(x_0)$ and generate new data points:
\begin{equation}
\begin{gathered}
    p_\theta(x_0) = \int p_\theta(x_{0:T})\,dx_{1:T},\\
    \text{ where }\;p_\theta(x_{0:T}) := p_\theta(x_T) \prod_{t=1}^T p_\theta(x_{t-1}|x_t).
\end{gathered}
\end{equation}
The learning objective is to find the parameters that maximize the variational lower bound on the expected log-likelihood:
\begin{multline}
        \mathds{E}_{q(x_0)}[\log p_\theta(x_0)] \geq
        \mathds{E}_{q(x_0,\ldots,x_T)}\left[\log \frac{p_\theta(x_{0:T})}{q(x_{1:T}|x_0)}\right].
\end{multline}
Ho et al.\ presented Denoising Diffusion Probabilistic Models (DDPMs) by specifying the forward diffusion process to use a fixed predefined Gaussian distribution with a variance schedule $\left\{\alpha_t\in (0,1]\right\}_{t=1}^T$~\cite{DDPM}, as follows:
\begin{equation} \label{eq:ddpm_q}
        q(x_{1:T}|x_0) = \prod_{t=1}^T q(x_t|x_{t-1}),
\end{equation}
\begin{equation*}
            \text{ with }q(x_t|x_{t-1}):=\mathcal{N} \left( x_{t-1}\sqrt{\dfrac{\alpha_t}{\alpha_{t-1}}}, \left( 1 - \sqrt{\dfrac{\alpha_t}{\alpha_{t-1}}}\right)I\right).
\end{equation*}
Given this, we can express the intermediate latent variables in a closed form:
\begin{equation}\label{eq:closedform}
    x_t = \sqrt{\alpha_t}x_0 + \sqrt{1 - \alpha_t} \epsilon, 
\end{equation}
which shows that $x_t$ is a linear combination of the initial point and a Gaussian noise variable $\epsilon \sim \mathcal{N}(0,I)$. Equation~\ref{eq:closedform} allows DDPM to reparameterize the learning objective to learn a predictor $\epsilon_\theta(x_t)$ of the noise to be removed to recover $x_0$.

Extending upon this, Denoising Diffusion Implicit Models (DDIM)~\cite{DDIM} propose to replace DDPM's Markov chain with a non-Markovian forward process, where each step depends also on the initial data point, enabling a deterministic and faster sampling mechanism compared to the original DDPM approach. The DDIM denoising process, which transforms an initial Gaussian noise vector $x_T$ to an image $x_0$ is given by:
\begin{equation}
    x^{(t)}_0 = \dfrac{x_t - \sqrt{1-\alpha_t}\epsilon_\theta(x_t)}{\sqrt{\alpha_t}}, 
\end{equation}
where $x^{(t)}_0$ denotes the estimation of $x_0$ from time step $t$. A single-step denoising process can thus be expressed as:
\begin{equation}
    x_{t-1} = \sqrt{\alpha_{t-1}}x^{(t)}_0 + \sqrt{1-\alpha_{t-1}}\epsilon_\theta(x_t).
\end{equation}
This is a recursive process starting from time step $T$: as $t$ decreases, $x^{(t)}_0$ becomes a more accurate estimate of $x_0$. Song et al.\ denote this recursive denoising process by $\mathcal{D}_\theta(x_T)$~\cite{DDIM}.

\subsection{Latent Diffusion}
Diffusion models are commonly implemented using a U-Net architecture~\cite{U-Net} that learns how to approximate the noise variable $\epsilon$.
However, training a full diffusion model directly on high-resolution images is computationally demanding.
To address this, a technique known as \emph{latent diffusion}~\cite{Rombach_2022_stable} is often used to compress the image into an intermediate latent representation, reducing memory consumption at training time.
Once the noise predictor is trained on this lower-dimensional space, the output of the diffusion model must be mapped back to the image space to generate an image. We denote $\mathcal{E}$ the compression function and $\mathcal{E}^{-1}$ the function that maps from latent to image space.
In practice, $\mathcal{E}$ and $\mathcal{E}^{-1}$ are often parameterized by a Variational Autoencoder (VAE). For example, the popular Stable Diffusion v2.1 uses AutoencoderKL~\cite{kingma2022autoencodingvariationalbayes}.
%For the remainder of this paper, we will denote this transformation function as $\mathcal{E}$.

%\mjm{we might not want to use the same notation for both this encoding and the attack's encoding, especially if we have experiments with different VAEs}.

% \jun{$\epsilon$ is used to refer to both the gaussian noise estimation and the transformation budget, we'll need to change it at some point.}
% \mjm{Yes, let's use $\delta$ for the transformation budget.}

\subsection{Digital Watermarking}
Digital watermarking emerged from the need to protect copyright as digital media became widespread and easy to distribute~\cite{tirkel1993electronic}.
Early methods considered an adversary attempting to remove watermarks to misappropriate the content. To address this threat model, watermarking techniques often rely on steganography to embed watermarks directly into the content---unlike digital signatures, which are typically appended in the content's metadata and can be stripped out. 
This inseparability from the content is a design goal of watermarking that distinguishes it from other authentication methods.

The rise of generative AI has introduced new use cases for watermarking. Research in this field has focused on the use of watermarks to protect the intellectual property of model owners~\cite{boenisch2021systematic} and dataset curators~\cite{du2024sok}. More recently, watermarking has been proposed as a means to reliably identify AI-generated content~\cite{zhao2024sok}. For image generation, such watermarking schemes aim to distinguish digital photographs and human-made art from artificially generated images.

An et al.~\cite{an2024waves} categorize image watermarking techniques into two main classes:

\para{Post-processing} techniques embed the watermark directly into the image. This includes traditional watermarking schemes based on signal-processing methods, such as those that encode the watermark in the image's frequency domain~\cite{cox2007digital,al2007combined}, as well as recent ML-based approaches that involve training models, such as autoencoders, to learn how to insert and subsequently extract an imperceptible watermark from an image~\cite{zhu2018hidden,zhang2019robust,tancik2020stegastamp}.
 
\para{In-processing} watermarking schemes are embedded within the generation process. This typically requires retraining the model~\cite{yu2021artificial,zeng2023securing,lukas2023ptw}, fine-tuning it~\cite{fernandez2023stable}, or modifying its sampling procedure. An approach specific to diffusion models, first proposed by \TR, is to embed the watermark in the model's latent space~\cite{wen2023treering,yang2024gaussian,gunn2024undetectable}. Latent-space watermarking has gained popularity due to its imperceptibility and the added difficulty it poses for adversaries: removal attacks require approximating the inversion of the diffusion process without knowledge of the model, as it is assumed to be proprietary.

\TR is a representative and meaningful target for study. As the first scheme to propose embedding the watermark in the model's latent space, \TR has been highly influential, with subsequent schemes closely following its embedding mechanism.
Given the growing adoption of latent-space watermarking, a detailed analysis of \TR has broader implications for the design and evaluation of future schemes.
For example, our use of VAEs to mount attacks in the latent space highlights a broader potential vulnerability applicable to other latent-space watermarking schemes.

\subsection{\TR Watermarking}
The \TR watermarking scheme, proposed by Wen et al.\ relies on one key property of DDIM models~\cite{wen2023treering}. Dhariwal et al.\ discovered that the DDIM sampling process can be inverted, mapping an image $x_0$ back towards the initial noise vector $x_T$~\cite{dhariwal}. Given a learned noise function $\epsilon_\theta$, we can re-estimate $x_T$ with:
\begin{equation}
    x_{t+1} = \sqrt{\alpha_{t+1}}x_0^{(t)} + \sqrt{1-\alpha_{t+1}}\epsilon_\theta(x_t).
\end{equation}
According to Dhariwal et al., this inversion process reasonably estimates the initial latent variable if two conditions hold: (i) the inversion process reintroduces noise in small steps, i.e., the difference between $x_{t}$ and $ x_{t+1}$ is small; and, (ii) $x_{t-1}-x_t \approx x_{t+1}-x_t$. If these two hold, the inversion process can accurately reconstruct the initial noise vector, enabling the key detection scheme of \TR watermarking. Wen et al.\ denote this inversion process by $\mathcal{D}^{\dagger}_\theta(x_0)$.

\begin{figure}[!ht]
    \centering
    \includegraphics[trim={0 2cm 0 0},clip,width=\linewidth]{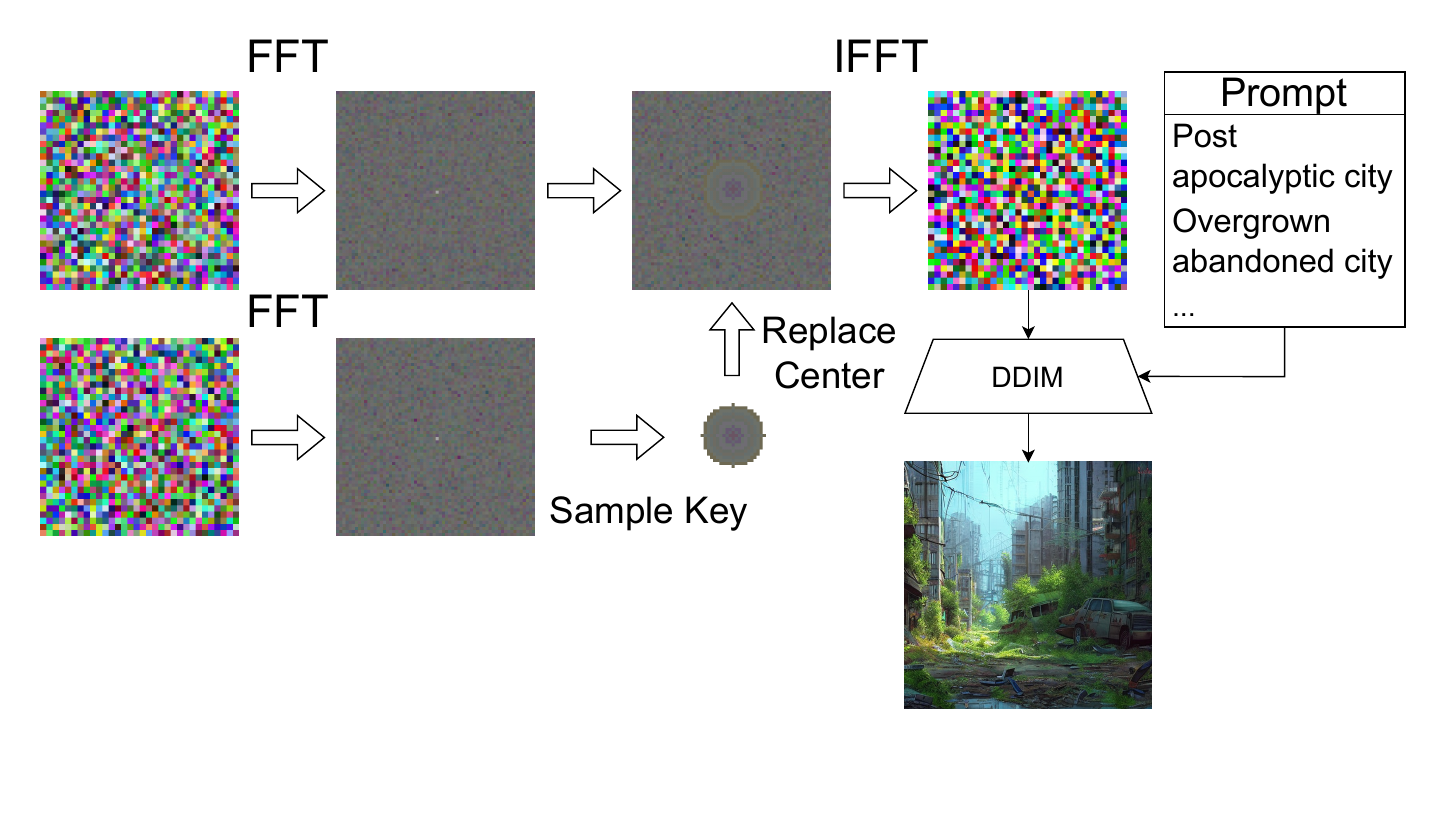}
    \vspace{-0.4cm}
    \caption{Embedding mechanism. The radius of the ring determines the number of pixels we sample from the frequency domain. The ring is embedded in the center of the initial Gaussian noise vector in Fourier space. After applying Inverse Fast Fourier Transform (IFFT), the resultant noise vector is fed to the  diffusion model for image generation.}
    \label{fig:treering_embed}
    \vspace{-0.5cm}
\end{figure}

\Cref{fig:treering_embed} illustrates the embedding process. A predefined key is encoded as a pattern in the central region of the Fourier space of the initial noise vector.
This region, which contains the low-frequency components, corresponds to global image characteristics. Embedding a watermark here diffuses its information across these high-level features, making the watermark not only more imperceptible but also more robust to image manipulations such as rotation and cropping. As the watermark becomes entangled with the image's semantic content, pixel-level operations that remove it inevitably destroy the image's meaning, rendering the image unusable from the adversary's perspective.
Wen et al.\ experimented with various key patterns and found that the most robust to rotations is a circular pattern composed of multiple rings, each ring with a constant value sampled from the frequency domain of a Gaussian vector. To generate a new image, the modified vector is transformed back to the latent domain via IFFT and denoised into an image with standard DDIM backward diffusion.

\begin{figure}[!ht]
    \centering
    \includegraphics[trim={0 0.4cm 0 0},clip,width=\linewidth]{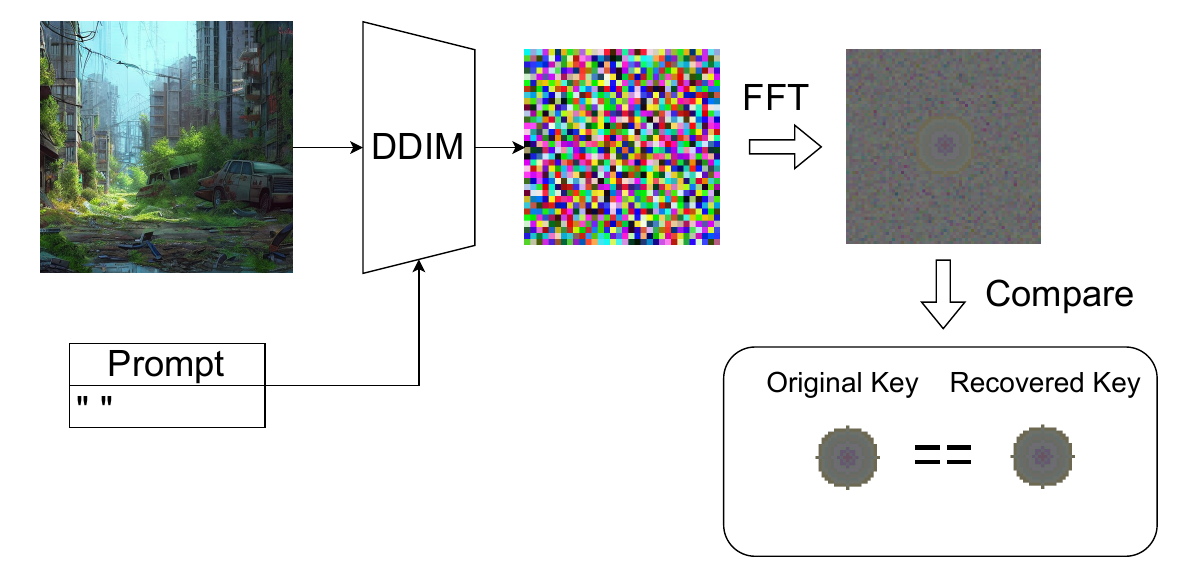}
    \vspace{-0.4cm}
    \caption{Detection scheme. Using forward diffusion, \TR estimates the initial noise vector used to generate the image with an empty prompt. It then extract the key in the center and thresholds its distance to the original key to make a final decision.}
    \label{fig:treering_detect}
\end{figure}

Detecting the watermark is more challenging, especially for conditional models that take a text prompt as input. For perfect reconstruction, both the output image and the original prompt are needed, as the prompt conditions the diffusion process to generate an image that aligns with the user's intent. However, for privacy reasons, model owners must allow users to delete their prompt history. Consequently, DDIM inversion without access to the original prompt can only approximate the initial noise vector (see \Cref{fig:treering_detect}).
That said, Wen et al.\ demonstrate that this approximation is sufficiently accurate to enable watermark detection even without the original prompt.

\subsection{Previous Attacks}
\TR's security has been evaluated in several studies, including the original \TR paper~\cite{wen2023treering}.% that overlooked the \emph{semantic} nature of \TR watermarks.
This evaluation showed that \TR is robust to rotation, cropping, rescaling, and JPEG compression~\cite{wen2023treering}.
However, these image transformations lack fine-grained control in the latent space necessary to target the watermark without substantially altering the image.
%Embedding the watermark into the latent space not only achieves high imperceptibility but also hinders removal attacks in the image space, as attackers must understand how image operations translate to the latent space.
%Wen et al.'s first evaluation showed that, indeed, \TR is robust to rotation, cropping, rescaling, and JPEG compression~\cite{wen2023treering}.
% and Gaussian blurring
%, as the attacks do not evade watermark detection without significantly decreasing image quality

%Wen et al.'s findings show that \TR is robust against rotation, JPEG compression, Gaussian noise addition, Gaussian blurring, and cropping+rescaling, as the attacks do not evade watermark detection without significantly decreasing image quality. For example, cropping+rescaling 70\% of the image and rotating it 45 degrees removes 30\% and 50\% of the original image, respectively. Since these loss margins destroy the semantics of the image, we consider them unacceptable in most adversarial use cases.

\smallskip
\noindent Next, we summarize follow-up works that have proposed more sophisticated attacks against \TR watermarking.

\para{Regeneration attacks.} 
Zhao et al.\ propose \emph{regeneration attacks}, which re-encode the image with an off-the-shelf encoder, inject noise, and subsequently decode to reconstruct the image~\cite{zhao2023invisible}.
The authors found that these attacks are effective against pixel-level watermarking schemes like Stable Signature, but they have limited success against \TR.
A more extensive evaluation by An et al.\ demonstrates that these attacks can be further tuned to increase their performance but then significantly degrade image quality~\cite{an2024waves}.

Zhao et al.\ instantiate the encoder--decoder pair of a regeneration attack with public auxiliary models, such as a VAE or a diffusion model. These attacks are closest to our attacks in that they rely on a public auxiliary model. However, our attack strategy is more precise, as our attacks exploit weaknesses specific to the \TR embedding mechanism, rather than indiscriminately adding noise to the latents.

\para{Adversarial attacks.} In the most extensive survey of watermark removal attacks to date, An et al.\ also explore \emph{adversarial attacks}, which are characterized by crafting adversarial examples against the watermark detector. The success rate of adversarial attacks is overall lower than that of regeneration attacks~\cite{an2024waves}, but they retain most of the original image, which is a desirable property for a removal attack.
An et al.\ distinguish between two types of adversarial attacks: \emph{surrogate detector} and \emph{embedding} adversarial attacks.

\para{Embedding attacks.} They rely on an existing encoder to map images to their corresponding latent representation. 
The adversary uses the encoder to find a small perturbation of the image that maximizes the distance between original and perturbed latents~\cite{an2024waves}. By maximizing this distance, the attacker aims to push the latent representation of the perturbed image outside the ``watermarked'' region of the latent space. An et al.\ evaluated embedding attacks extensively and showed that they are only effective if the attack's encoder exactly matches the one used by the victim's diffusion model.
Although this attack could be improved, embedding attacks do not target specific parts of the embedding but rather randomly shift the entire embedding, causing significant changes in the final image.

\para{Surrogate detector attacks.} The adversary first trains a surrogate classifier to detect the presence of a watermark, mimicking the behavior of the actual watermark detector. Next, they aim to find a small perturbation in the image that causes the surrogate detector to misclassify it, in the hope that this misclassification will transfer to the watermark detector.
Unlike embedding attacks, these attacks do not explicitly assume access to the victim's encoder; however, gathering the data to train the surrogate can be challenging in practice.

An et al.\ explore various methods to obtain training data for the surrogate detector. One of the training procedures requires white-box access to the victim's diffusion model to generate a dataset of image pairs, each pair consisting of a watermarked image and its non-watermarked counterpart.
This is an unrealistic assumption: it implies full access to the victim's model, eliminating the need for an attack, as the adversary could directly sample from the model to generate non-watermarked images. A more realistic setting is to train the surrogate classifier with \TR images and an auxiliary publicly-available image dataset. However, this attack did not evade the \TR detector~\cite{an2024waves}.
An et al.\ hypothesize that the reason is the difference between the public dataset distribution and that of the generated images, which may have led to spurious features that are independent of the watermarks~\cite{an2024waves}.

% \para{Adaptive attacks.} Lukas et al.\ propose \emph{adaptive attacks}~\cite{lukas2024leveraging}, which can be seen as a particular instance of embedding attacks~\cite{an2024waves}. Adaptive attacks assume access to the key generation procedure of the watermarking scheme. During the preparation of the attack, the adversary generates a set of keys to fine-tune a less capable version of the victim's model. To deploy the attack, the adversary can use the resulting fine-tuned model in two ways: \emph{adversarial noising}, where the adversary finds an adversarial example through a perturbation that maximizes the distance between the decoded keys; and \emph{adversarial compression}, which has the same optimization objective as adversarial noising, except that the perturbation is in the form of a compression of the image that minimizes the LPIPS loss.
% Adaptive attacks assume knowledge of the watermarking scheme, e.g., the key generation process, and are therefore less practical than other adversarial attacks~\cite{an2024waves}.

\begin{figure*}[!ht]
    \centering
        \subfloat[$t=40$]{
            \centering
            \includegraphics[width=0.18\textwidth]{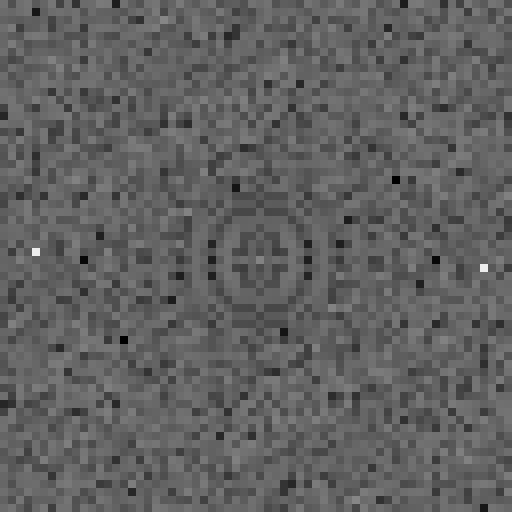}
        } 
    \hfill
        \subfloat[$t=30$]{
            \centering
            \includegraphics[width=0.18\textwidth]{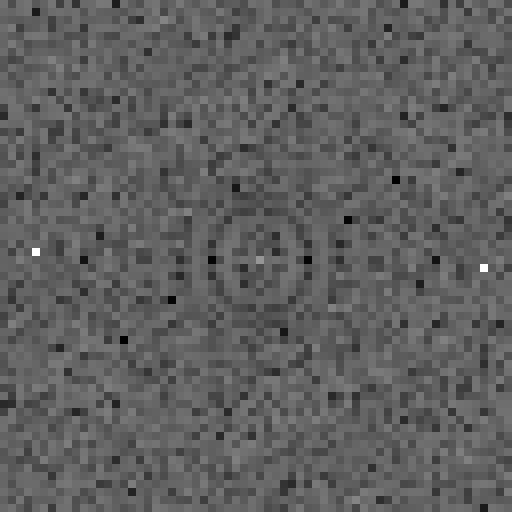}
        }
    \hfill
        \subfloat[$t=20$]{
            \centering
            \includegraphics[width=0.18\textwidth]{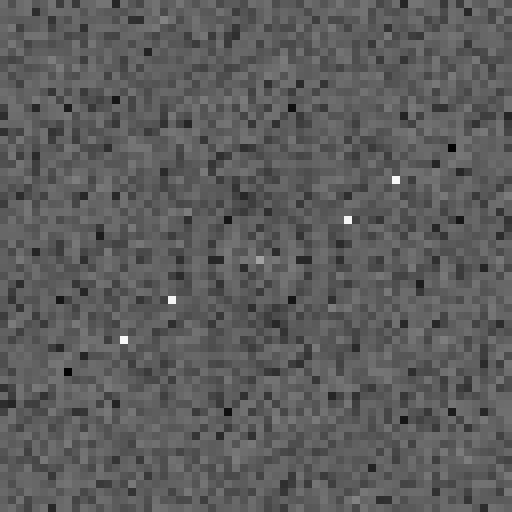}
        }
    \hfill
        \subfloat[$t=10$]{
            \centering
            \includegraphics[width=0.18\textwidth]{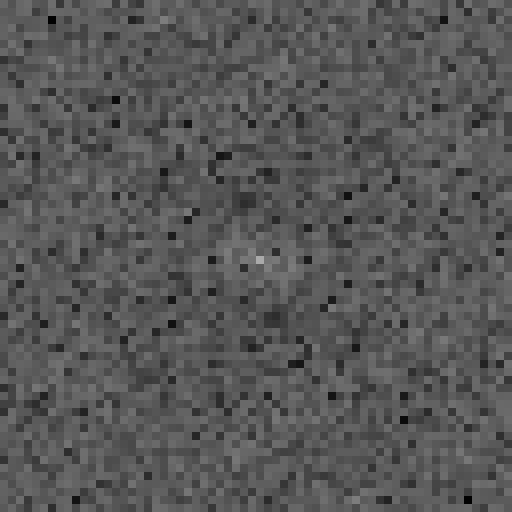}
        }
    \hfill
        \subfloat[$t=0$]{
            \centering
            \includegraphics[width=0.18\textwidth]{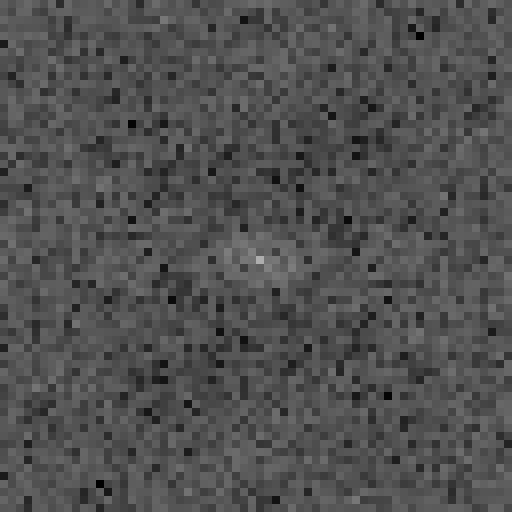}
        }
    \\
        \subfloat[$t=40$]{
            \centering
            \includegraphics[width=0.18\textwidth]{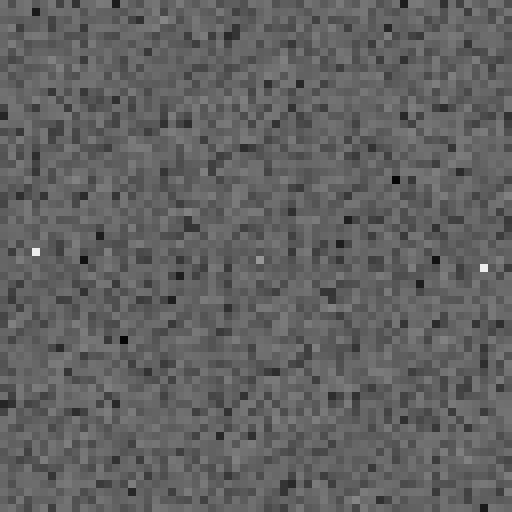}
        } 
    \hfill
        \subfloat[$t=30$]{
            \centering
            \includegraphics[width=0.18\textwidth]{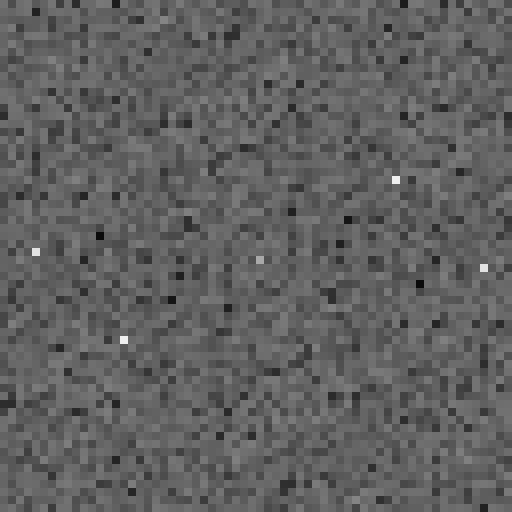}
        }
    \hfill
        \subfloat[$t=20$]{
            \centering
            \includegraphics[width=0.18\textwidth]{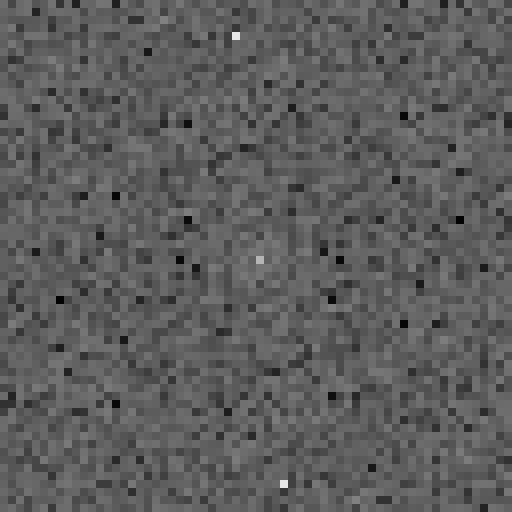}
        }
    \hfill
        \subfloat[$t=10$]{
            \centering
            \includegraphics[width=0.18\textwidth]{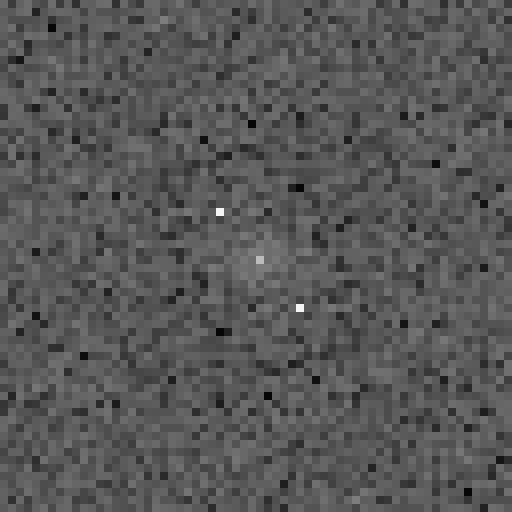}
        }
    \hfill
        \subfloat[$t=0$]{
            \centering
            \includegraphics[width=0.18\textwidth]{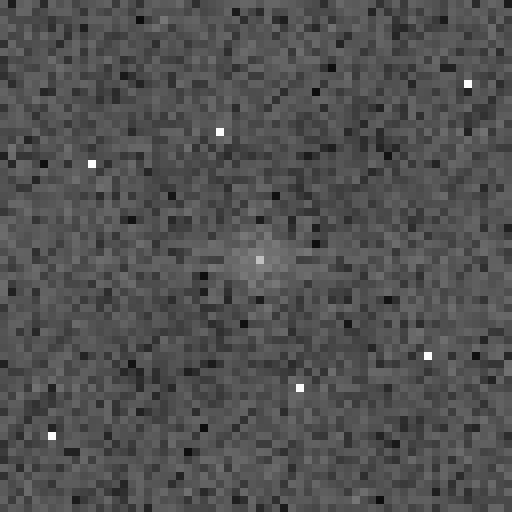}
        }
\caption{The Fourier transform (magnitude) of the fourth channel for the intermediate latents of the backward diffusion process. The first row of figures, (a)--(e), are the intermediate latents of an image with the watermark embedded. The second row, (f)--(j), are the intermediate latents generated with the same prompt but excluding the watermark. The plots are all in log scale and centered around the zero-frequency component. We omit the axes for emphasis.}
\vspace{-.2cm}
\label{fig:w_w/o_latent}
\end{figure*}

\para{Adaptive attacks.} Lukas et al.\ propose \emph{adaptive attacks}~\cite{lukas2024leveraging}, which can be seen as a combination of embedding and surrogate detector attacks~\cite{an2024waves}. The instantiation of these attacks for \TR relies on a less capable version of the victim's model. To deploy the attack, the adversary can use this surrogate diffusion model in two ways: \emph{adversarial noising}, where the adversary finds an adversarial example through a perturbation that maximizes the distance between decoded keys (embedding attack); and \emph{adversarial compression}, which compresses the image while minimizing the LPIPS loss.
Lukas et al.\ evaluate these attacks assuming that the adversary has a diffusion model with the same architecture as the victim's.
This is a strong assumption: if the victim's model is proprietary, the adversary is unlikely to have access to a diffusion model with the same architecture, and it is unclear how much attack performance degrades if this assumption does not hold.

\smallskip
We present novel adversarial attacks with fewer assumptions on the adversary's access and knowledge. In particular, we use an auxiliary model, such as a public VAE, to formulate An et al.'s surrogate detector attacks in the latent space, eliminating the need for a dataset of non-watermarked images.
%that assume only query access to the victim's diffusion model and use an auxiliary publicly available dataset of non-watermarked images to train the surrogate model.

\section{Attack Methods}
In this section, we define our threat model and describe our attack methods in detail.

\subsection{Threat Model}\label{sect:threat_model}
We follow the threat model in previous watermarking studies, considering two parties: the \emph{model owner} and the \emph{adversary}. The model owner maintains a proprietary diffusion model and monetizes it by offering subscription-based access to prompt it (e.g., through an API).
Upon receiving a prompt, the model owner uses \TR to embed a \emph{key}, $k$, and generate a watermarked image.
We only consider proprietary models because \TR watermarks are embedded during generation and thus cannot be enforced on open-source models.

\para{Adversary's goal.}
The adversary's goal is to use the images generated by the hosted model for their own purposes without revealing that they were generated with \emph{that} model. To achieve that, the adversary's strategy is to apply an image transformation such that the watermark is no longer detected.

Although certain images might be more susceptible to attacks than others, the attacker's aim is to identify a general transformation that succeeds for an arbitrary image, implying ineffectiveness of the watermarking scheme.
To be more precise, the adversary's objective is to find an image transformation $\mathcal{F}$ such that, for any image $x$ generated with the queried model, it satisfies two conditions: 
\begin{enumerate}[label=\roman*.]
  \item $\left\lVert k-\tilde{k}\right\rVert_1>\tau$, where $\tilde{k}$ is \TR's approximation of the original key $k$ and $\tau$ is the decision threshold, and
  \item $\left\lVert\mathcal F(x)-x\right\rVert_{\infty}\leq\delta$, for a small $\delta>0$.
\end{enumerate}

%\noindent where $x^*:=\mathcal F(x)$ is the transformed image.

Although the maximum norm is a common metric to bound an image perturbation, the specific metric will depend on the adversary's use case and their tolerance to changes in the image. In \Cref{sec:evaluation}, we quantify how our attacks achieve the above objectives by measuring the decrease in detector performance under multiple common image similarity metrics.

\para{Adversary's capabilities.}
Since the target model is proprietary, the adversary has only black-box access: they can query it but have no knowledge of its parameters. Specifically, they observe only the output image and the input prompt that produced it. Additionally, the adversary is assumed to have limited computational resources; if they had sufficient resources, they could train a diffusion model from scratch instead of resorting to an attack. We therefore define a watermarking scheme as \emph{robust} if such a computationally bounded adversary cannot successfully remove the watermark without spending more resources than would be required to collect data and train a comparable diffusion model. However, the adversary can still leverage publicly available models, such as pretrained autoencoders, readily accessible on model-sharing platforms, such as Civitai~\cite{civitai} and Hugging Face~\cite{huggingface}.

%While this simplifies the training process of models, the availability of these auxiliary models facilitates our attacks.
%help to establish a set of tools and pipelines that are followed to produce text-to-image models. Due to the simplification of this process, there is a lack of consideration in security for text-to-image models, which allows us to mount our attacks.
% This is realistic because, in practice, a pretrained autoencoder is often used to compress input data into a latent space where the diffusion process is applied, training diffusion models more efficiently~\cite{Rombach_2022_stable}.

\para{Justification of the public VAE access assumption.}
Auxiliary models suitable for use in attacks are publicly available on model-sharing platforms. These platforms are driven by the need to make diffusion model development---such as training and fine-tuning---more cost-efficient and accessible.
Since training diffusion models from scratch is computationally expensive, practitioners often adopt latent diffusion, which reduces training costs by leveraging a pre-trained VAE. As a result, these VAEs are frequently reused and redistributed without modification.

Another common practice is to fine-tune existing diffusion models rather than retraining the full pipeline. For example, the most popular way to fine-tune a diffusion model is to apply LoRA methods~\cite{hu2021loralowrankadaptationlarge} to fine-tune only the U-Net component, while leaving the VAE unchanged~\cite{ruiz2022dreambooth}. Consequently, even if the fine-tuned model is not published, the original VAE would remain public. We observe numerous real-world instances of these practices. For example, Civitai~\cite{civitai}, a popular model-sharing platform, hosts a wide range of fine-tuned diffusion models that either retain the original VAE or continue training from publicly available checkpoints.

There is a plethora of examples of commercial diffusion models that rely on a VAE (see \cite{flux,dalle-vae,shuttleai,openai_sora,craiyon,clipdrop,leonardoai,dreamstudio,aiseoart,davinci_sdxl,cogview4,deepdreamgenerator}), and the trend of publishing the VAE used during training is also common in the industry. For instance, commercial text-to-image APIs such as those offered by Black Forest Labs~\cite{flux}, ShuttleAI~\cite{shuttleai}, and BigModel~\cite{cogview4} provide access to their VAEs. Even OpenAI has published the VAE for its text-to-image model, DALL-E~\cite{dalle-vae}. These practices make access to the VAE associated with the target diffusion model a realistic and practical assumption. Moreover, even if the exact VAE is not publicly available, the abundance of VAEs trained for other diffusion models makes it likely that a suitable alternative is available.

\subsection{Flaws of \TR Watermarking}~\label{sec:flaws}
The high accuracy of \TR's detector despite the large amount of noise added in Wen et al.'s evaluation suggests that the approximation of the original initial latent is remarkably distinct from its non-watermarked counterpart.
This inherent difference between watermarked and non-watermarked latents exposes a vulnerability in \TR that can be exploited.

\para{Non-Gaussian nature of watermark latents.}
The embedding mechanism of \TR violates the assumption that the initial latent follows a Gaussian distribution. 
Recall that \TR draws both the initial latent vector and the values of the key ring from a Gaussian distribution.
However, that does not imply that the combination of the two together will be normally distributed.

To demonstrate the difference between the watermarked and non-watermarked latent distributions, we examine the Fourier transform of the channel in which the key is embedded. \Cref{fig:w_w/o_latent} shows the progression of the intermediate latent through the backward diffusion process, from the step when the key is embedded ($t=40$) to the final latent ($t=0$). As shown in the figure, a ring structure is clearly visible in the first diffusion steps and is gradually removed after each step until, in the final diffusion step ($t=0$), the watermark is almost imperceptible. However, remnants of the original key structure remain visible at the center of the latent representations.
%upon closer inspection of the figure, we still notice a ring structure at the center of the latent representations: these are remnants of the original key.

After applying backward diffusion, images are obtained by applying a transformation function parameterized by a neural network (e.g., a VAE) that has been pretrained to generate images. In the \TR scheme, to detect the watermark, the image needs to be re-encoded back to latent space before forward diffusion can be used to recover the key. Although this process incurs a small amount of information loss, for key detection to work, this step must be sufficiently accurate to approximate the initial latent vector, suggesting that traces of the key must prevail to some extent.

This observation is the basis of our attacks: the difference in the distributions of the watermaked and non-watermarked latents naturally suggests attack strategies that aim to bring the watermarked latents back to their original distribution.

\begin{figure}
    \centering

\includegraphics[width=\columnwidth]{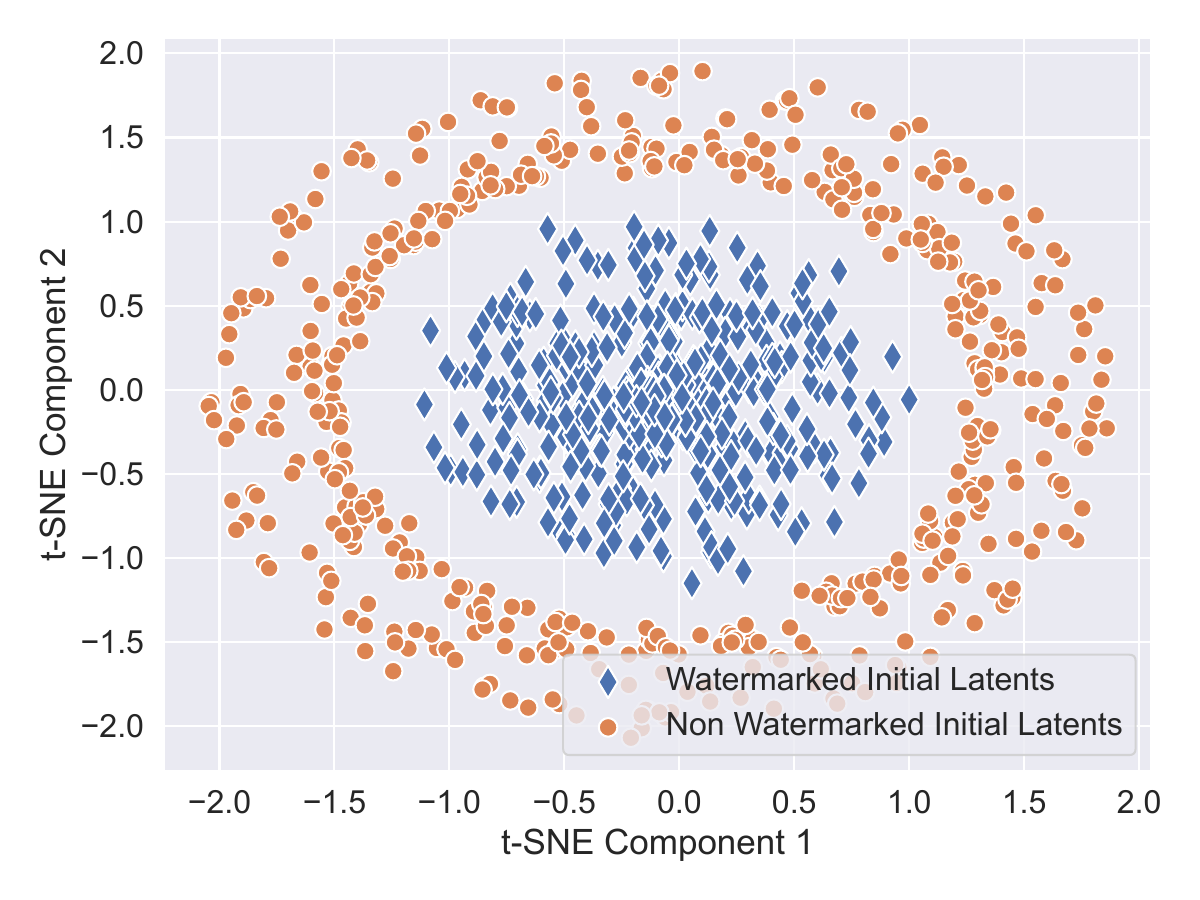}
    \vspace{-.75cm}
    \caption{Two t-SNE components of the initial latents of the Wm \& UnWm dataset. Data points for watermarked images are tightly clustered together around the center. A Gaussian kernel could easily separate the two classes.
    }
    \label{fig:tsne_init_lats}
    \vspace{-.4cm}
\end{figure}

\para{Separability of the latents.}
To further demonstrate the difference between the watermarked and non-watermarked latent distributions, we perform dimensionality reduction and plot two t-SNE (t-distributed stochastic neighbor embedding) components of the initial latents of a set of watermarked images and their non-watermarked counterparts (from the dataset ``Wm \& UnWm'' described in \Cref{sect:datasets}). %We used \TR with Stable Diffusion v2.1 and stable diffusion prompts wmvsunwm.
As \Cref{fig:tsne_init_lats} confirms, the two classes are perfectly separable, suggesting that the initial watermarked latent variables form a hypersphere in the latent space. Moreover, \Cref{fig:tsne_init_lats} shows this hypersphere exhibits compactness and well-defined boundaries.

This hints on the type of image transformation $\mathcal F$ that would effectively bypass the detector of a \TR scheme. Such transformations require modifying the latent representation such that the forward diffusion process $\mathcal{D}^{\dagger}_\theta(x_0)$ does not map the image back to the original watermarked latent hypersphere. %(projected in \Cref{fig:tsne_init_lats}). 
To enable such an attack, the adversary should focus on mapping the image back into latent space rather than defining transformations in the image space.
Given the non-Gaussian nature of the intermediate latents, it may suffice to recover an intermediate latent.
Such mapping can be achieved by using autoencoders, e.g., Stable Diffusion's VAE~\cite{Rombach_2022_stable}, SDXL's VAE~\cite{podell2023sdxl} and other models with similar architectures~\cite{Ostris}.
% Simple transformations like rotation and crop+rescale~\cite{an2024waves} cause considerable shifts in image space because they affect the position of all pixels within the image.
% However, they are not disruptive enough in the latent space to derail the inversion process toward the non-watermarked region.

\subsection{Suitability of a Surrogate Detector Attack}
Assuming we have the means to recover the latent space representation of an image, we ask the question of which attack strategy is most effective in pushing a point out of the watermarked hypersphere.

Regeneration attacks produce $x^*_0$ by adding Gaussian noise to $\mathcal{E}(x_0)$, where $\mathcal E$ encodes an image back to latent space.
This noise completely disrupts the latent representation $\mathcal{E}(x_0)$ and changes the output image distribution. However, because the noise indiscriminately shifts the entire latent representation, the quality of the image suffers as a result.

Adversarial attacks aim to shift the latent via solving an optimization problem. For example, embedding attacks directly solve $\max_{x^*_0}\lVert \mathcal{E}(x^*_0) - \mathcal{E}(x_0) \rVert$ such that $\lVert x^*_0 - x_0 \rVert < \delta$, where $x^*_0:=\mathcal F(x_0)$ is the transformed image.
% using Projected Gradient Descent (PGD). 
Although more refined than a regeneration attack, this strategy is not optimal. Due to the non-Gaussian nature of the watermark's latent distribution, the overall goal of the transformation $\mathcal F$ should not be to merely maximize the perturbation's effect on the latent, as formulated in the optimization objective. Instead, the objective should also ensure that the latents of the transformed images are distributed like the original latents, i.e., $\mathcal{D}_\theta^\dagger (x^*_0) \sim \mathcal{N}(0;I)$, where $\mathcal{D}_\theta^\dagger(x^*_0)$ is the initial noise vector resulting from forward diffusion.%, i.e., the transformation should be distributed like the original latent distribution.

Surrogate detector attacks are better suited to achieve that. Training the surrogate detector is akin to modeling the watermarked hypersphere. In essence, such a surrogate detector is an approximation of the watermark detector independent of the implementation details of the watermark detection scheme. Thus, the loss of the surrogate detector provides a mechanism to mount an adversarial attack with precisely the desired objective: after training the surrogate detector, the adversary solves $\min_{x^*_0} L(\mathcal{E}(x^*_0),y_{\text{target}})$ s.t.\ $\lVert x^*_0 - x_0 \rVert < \delta$, where $L$ is the loss of the surrogate detector.

This is a different objective from the surrogate detector attack evaluated by An et al.~\cite{an2024waves}. Their surrogate detector was trained on watermarked images, but, because it is unrealistic to assume the adversary can generate the non-watermarked versions of those images, they used publicly available images. Public datasets follow a different distribution than the model's output distribution and, as a result, An et al.'s surrogate detector picked up on features that might have helped to distinguish between the two image distributions, but that were independent of the watermark, leading to an inadequate surrogate for finding adversarial examples.

Instead, we propose training the surrogate detector with the latent representation of the images, increasing the likelihood that it discriminates based on features dependent on the watermark.
Our surrogate detector is therefore more likely to capture the watermarked region and generalize better than An et al.'s.
This encoding step is crucial because the success of a surrogate detector attack hinges on how well the surrogate detector approximates the \TR detector.

\begin{figure}[!t]
    \centering
    \includegraphics[width=\linewidth]{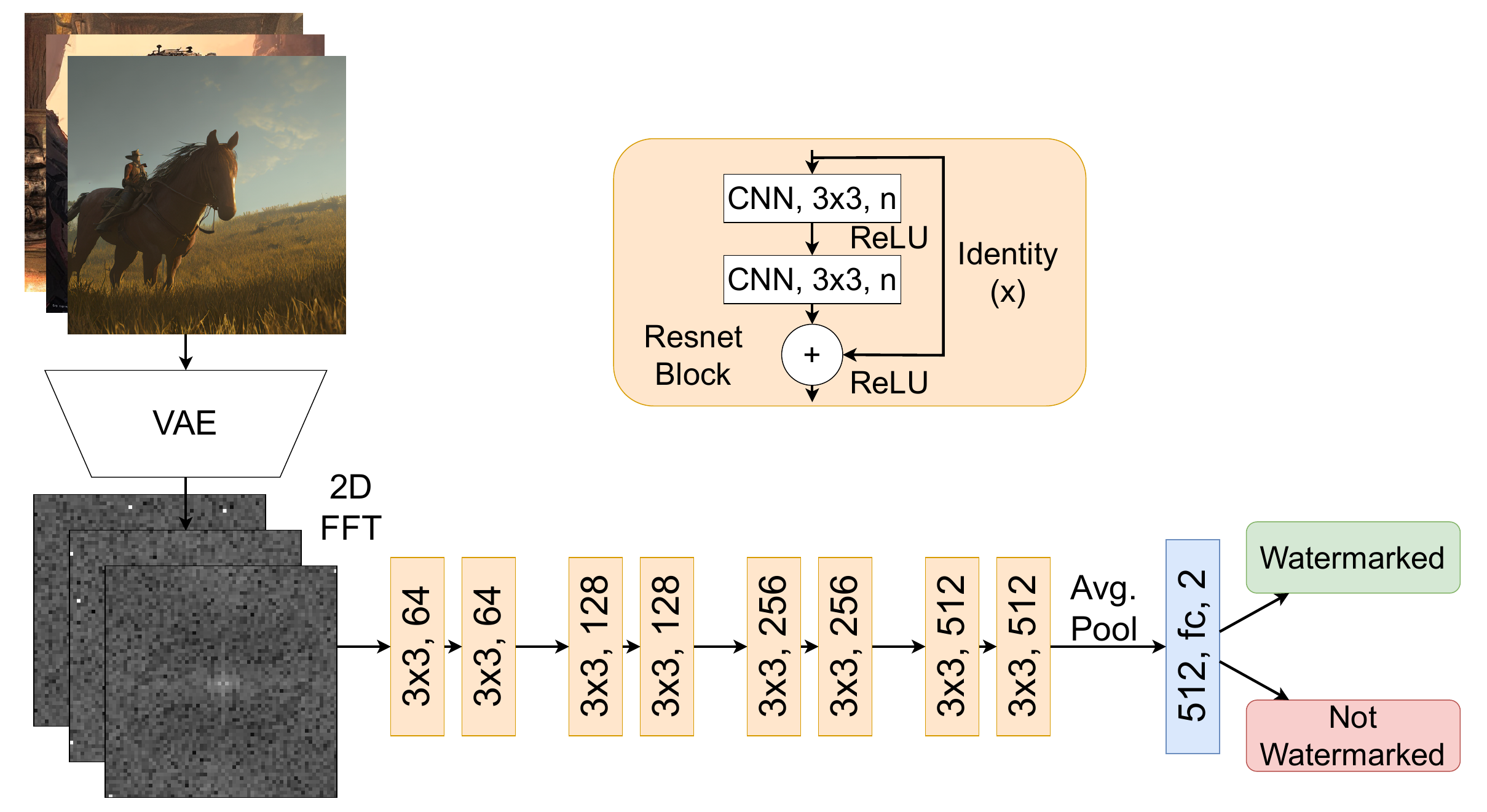}
    \caption{Surrogate detector training pipeline. The input data is a mixed batch of watermarked images generated via Stable Diffusion and either non-watermarked images generated via the same diffusion model or public images from ImageNet.}
    \label{fig:surrogate_pipeline}
\end{figure}

\begin{figure}[!t]
    \centering
    \includegraphics[width=\linewidth]{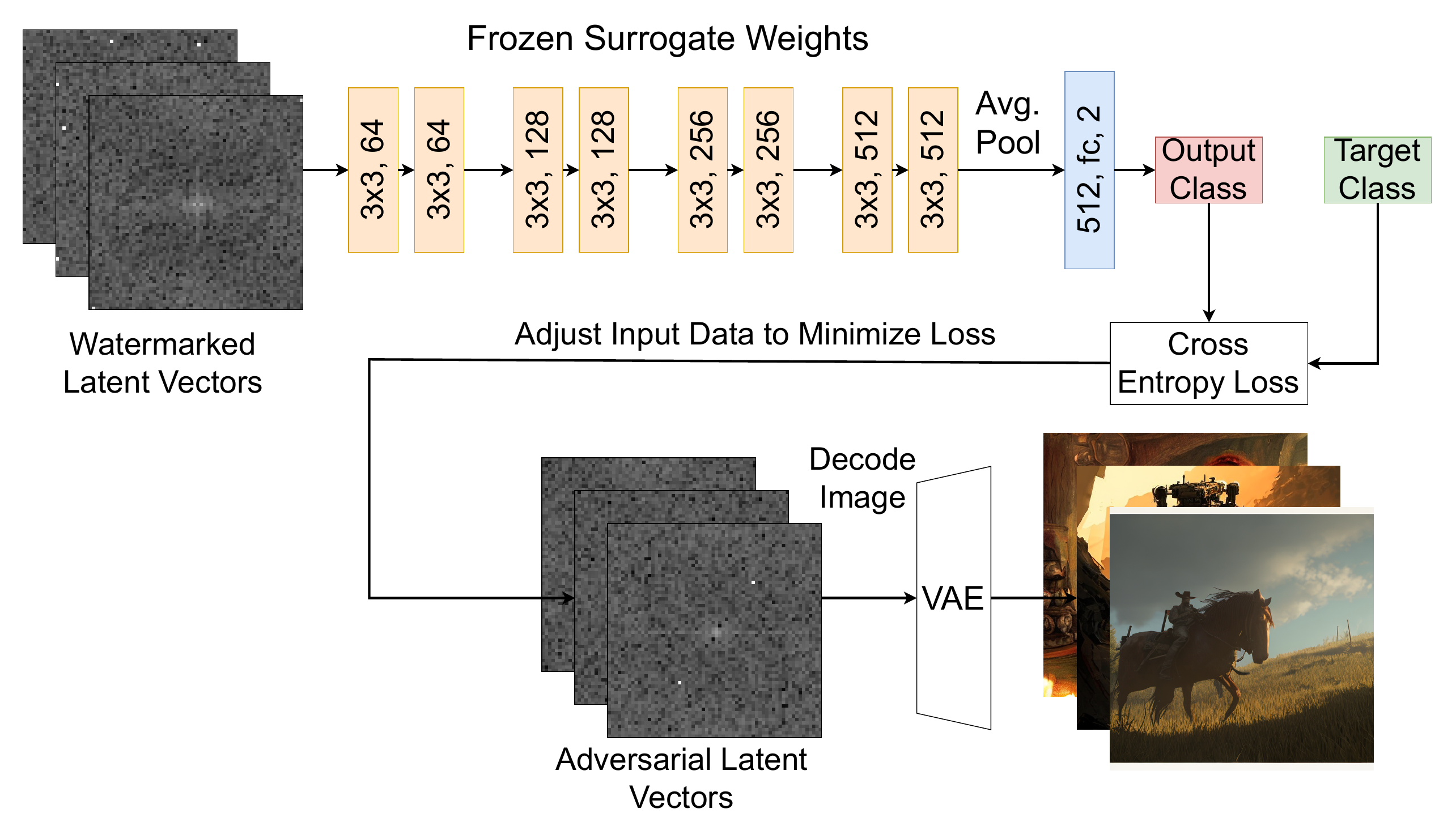}
    \vspace{-0.5cm}
    \caption{PGD attack pipeline. The attack uses the surrogate detector trained as shown in \Cref{fig:surrogate_pipeline} to alter the latent representation of the attacked image via Projected Gradient Descent (PGD). The target class is the ``non-watermarked'' class.}
    \label{fig:pgd_pipeline}
\end{figure}
%\mjm{there are some inconsistencies throughout the paper:
%- sometimes we specify that the target model is a DDIM and others we keep it general and say diffusion model.
%- Encoder/VAE/encoder--decoder/diffusion model
%- Embedding/latent
%- Hidden message/watermark/key
%we need to clarify the usage of these terms and be consistent 
%}

\subsection{Overview of Our VAE-Based Attacks}

Our attack implementation is divided into three main components: the \emph{surrogate detector model}, the \emph{autoencoder model}, and the \emph{Project Gradient Descent (PGD) attack}. In this section, we describe how the adversary trains these models and combines them to deploy the attack.

\para{Attack preparation.}
The first step of the attack is to collect a dataset of watermarked vs.\ non-watermarked images. Since the adversary has query access to the victim's model, they can simply query it to generate the watermarked images. 
To obtain non-watermarked images, the adversary will draw them from public image datasets, ideally as semantically similar as possible to the generated images, as defined by the prompt.

To encourage the surrogate detector to learn to discriminate watermarked vs.\ non-watermarked images, the adversary uses a pretrained autoencoder that maps the images to latent space.
In our evaluation, we start by assuming that this autoencoder is identical to the one of the victim and, then, evaluate the setting in which these two autoencoders are different.

\para{Surrogate detector model.}
Our surrogate detector model is based on Resnet18~\cite{he2015deepresiduallearningimage}, a common neural network architecture in image classification tasks. Since our input data is in the frequency domain, we adapted its design to handle the frequency's magnitude and phase as a complex number. A minor implementation detail is that, due to this representation, the output is also a complex number which is not fully supported by Pytorch cross-entropy loss. As such, we only use the real components of the output likelihood estimates when we train our surrogates on latent representations.

The surrogate is thus a binary classifier whose output is whether or not an image is watermarked based on its input data representation (see \Cref{fig:surrogate_pipeline}). We use a batch size of 32, Adam optimizer with a learning rate of $10^{-3}$, and optimize for cross-entropy loss. For model selection, we use a 7:3 evaluation split ratio, and save the model with the highest validation accuracy every 100 epochs for later use in the PGD attack.

\para{PGD attack.}
The ultimate goal of the attack is to construct an image close to the original target image that our surrogate detector misclassifies as ``non-watermarked.''
Our surrogate detector predicts the class of the image and calculates the cross-entropy loss between the output class and the target class, i.e., the non-watermarked class (see \Cref{fig:pgd_pipeline}). Backpropagation provides the gradient step required to shift our data towards being misclassified.
As shown in \Cref{algo:pgdattack}, we repeat this process for $N$ steps and introduce a step size parameter $\alpha$ to control the smoothness of the perturbations. We take $N=200$ and $\alpha=0.05$, a setup similar to the one by An et al.~\cite{an2024waves}.

\begin{algorithm}[t]
    \caption{PGD Attack}\label{algo:pgdattack}
    \begin{algorithmic}[1]
    \Require Surrogate Model $\theta_C$, budget $\delta$, image $x$, VAE $\mathcal{E}$, $N$ steps, target class $y$
    \State $\theta_{x_0} \gets \mathcal{E}(x_0)$ \Comment{Encode to latent space}
    \For{$n$ to $N$}
        \State $\hat{y} \gets \theta_C(FFT(\theta_{x_0}))$ \Comment{Fourier Transform (FFT)}
        \State $g_{\theta_{x_0}} \gets \nabla_{\theta_{x_0}} \text{BCELoss}(\hat{y},y)$
        \State $g_{\theta_{x_0}} \gets \mathcal{P}_\delta(g_{\theta_{x_0}})$ \Comment{Clip gradients to $\delta$ budget}
        \State $\theta_{x_0} \gets \theta_{x_0} - \text{Adam}(\theta_{x_0},g_{\theta_{x_0}})$
    \EndFor
    \State $x_0^* \gets \mathcal{E}^{-1}(\theta_{x_0})$ \Comment{Decode to image space}
    \State \Return $x_0^*$
    \end{algorithmic}
\end{algorithm}

An et al.\ use a perturbation budget to limit the amount of pixel changes.
Their budget is defined in the image domain, however, since our attack acts in the latent space, we must rescale the budget. To find the appropriate scale, we take the maximum number of watermarked latents in our dataset $p$ and set our perturbation budget to $\frac{1}{p}$. After completing the $N$ steps, we decode the latent into an image using the autoencoder.

%\smallskip
To sum up, once we have trained the surrogate, the deployment of the attack on a watermarked image requires encoding the image using the autoencoder, obtaining its Fourier transform, and performing the PGD attack with the loss of the surrogate on the image. The attacker can repeat the attack with different values of $\delta$ until the adversarial image on the surrogate transfers to the \TR detector.

\section{Evaluation Methodology}
%This section describes the methodology used to evaluate the attacks.

\subsection{Models}
We take Stable Diffusion v2.1~\cite{Rombach_2022_stable} as the victim's diffusion model. This version of Stable Diffusion uses AutoencoderKL~\cite{kingma2022autoencodingvariationalbayes} by default as its VAE to generate the image from the intermediate latent space. We use this VAE to encode and decode images for our surrogate detector attack. However, in \Cref{sect:abblation}, we also challenge the assumption of having the exact same VAE by evaluating the attack with different VAEs.

\subsection{Datasets} \label{sect:datasets}
To assess the effectiveness of our attacks, we used the ImageNet1K dataset~\cite{ILSVRC15}. Specifically, following the methodology established in previous work~\cite{an2024waves}, we used 7,000 images from the ImageNet1K validation dataset. Each image has a set of annotated bounding boxes around objects each corresponding to one of the classes in ImageNet1K.
We sample 7 images from each class and preprocess the image by cropping to the first annotated bounding box. Afterwards, we resize each image to 512x512 to match the output dimensions of Stable Diffusion. To match the classes from ImageNet1K, we queried Stable Diffusion v2.1 to generate an image for each class using the prompt: ``A photo of a  \texttt{<ImageNet Class>}.'' Each generated image was watermarked with the \TR scheme, and 7 images were generated for each class, each with a different random seed to introduce diversity in the outputs.

To train the surrogate detector, we collect two datasets:

% \mjm{renamed Real to Pub (Public). @Junhua: can you plot the graphs in Figures 8-10 to reflect this change?}

\para{Wm \& UnWm} represents the setting where the adversary can use the diffusion model to generate both the watermarked and non-watermarked images. It includes a total of 3,000 images: 1,500 images are generated using a victim diffusion model with \TR watermarks (Wm); the remaining 1,500 are generated using the same diffusion model, random seed, and image prompt, but skipping the watermarking step (UnWm).

\para{Wm \& Pub} caputres the more realistic setting where the adversary uses images in a public dataset as the non-watermarked training examples. This dataset has a total of 14,000 images, 7,000 images are generated using a victim diffusion model with \TR watermarks, the other 7,000 are preprocessed ImageNet images.

\smallskip
\noindent The Wm \& UnWm dataset is smaller since training an accurate surrogate detector with the non-watermarked version of the watermarked images requires significantly less data.

%To enable surrogate training on latent representations, we also needed to project our images to their latent representations. Using an autoencoder (AutoencoderKL by default), we project all images to latent space.
%These latents were cached for later evaluation of the attacks.

\subsection{Performance Metrics} \label{sect:perf_metrics}

In line with existing evaluations of watermarking techniques for generative models~\cite{an2024waves,wen2023treering,zhao2023invisible}, we have selected metrics to assess various aspects of the attack. These metrics measure the fulfillment of the two adversarial goals outlined in \Cref{sect:threat_model}: (i) the first set of metrics quantify the success in evading detection measured by the decrease in \emph{predictive performance} of the watermark detector after deploying the attack; (ii) the second set of metrics measure the impact of the attack on the target image by quantifying the degradation of the attacked image compared to the original.

As pointed out by An et al., these objectives are context-dependent and are subject to the adversary's use case. For this reason, we use multiple metrics for each objective, providing a comprehensive evaluation and a discussion of their practical implications. The two sets of metrics are as follows.

\subsubsection{Predictive Performance Metrics}

\noindent We assume that the reader is familiar with basic binary classification predictive performance metrics, such as Accuracy, Precision, Recall (TPR), and FPR, and we will only describe metrics that are less known or that are particularly common in the evaluation of watermarking techniques.

\para{TPR@1\%FPR} is the detector's TPR at a threshold chosen to ensure a 1\% FPR. A 1\% FPR threshold is commonly used in prior evaluations of watermarking schemes and is considered a stringent criterion~\cite{an2024waves,wen2023treering}.
%establish a strict criterion at which the watermark detector has a low probability of misclassifying a non-watermarked image as watermarked. A successful detector should be able to ensure a high TPR, while 

%This prioritizes the correct detection of legitimate non-watermarked images, while the TPR assesses the detection algorithm's ability to correctly identify watermarked images given these constraints.

\para{ROC-AUC} measures the \emph{Area Under the Curve} (AUC) of the \emph{Receiver Operating Characteristic} (ROC) curve, a common performance metric for binary classifiers. In this case, it quantifies the detector's ability to detect watermarked images correctly as an aggregate measure of the
detector's TPR--FPR trade-off for all possible decision thresholds.

\para{PR-AUC} is the Area Under the Precision--Recall curve. Like the ROC-AUC, this is a composite metric that measures the classifier's ability to ensure both high Precision and Recall (TPR), aggregating the classifier's trade-offs at all possible decision thresholds. 

We measure these metrics using the setup used in \TR's original paper~\cite{wen2023treering}, setting a 1:1 ratio between the two classes. For example, for the Wm \& UnWm dataset, we have 1,500 attacked watermarked images vs.\ 1,500 non-watermarked images. In \Cref{sec:base_rate}, we evaluate performance while varying this ratio.

\subsubsection{Image Degradation Metrics}
% \jun{All the score I reported in this part use the difference between the watermarked vs reference and attacked image vs reference quality measures, I would either need to re-run to get all the quality metrics again or state that here because a few of them required the use of reference images.}
% \mjm{I see. I brought back the difference metric and rephrased it a bit (please, check)}

\noindent We evaluate the visual impact of the attacks using several metrics, each reflecting different aspects of image quality.

\para{CLIP Score} is the cosine similarity between an image embedding and a textual embedding given a fixed text-to-image model. We use it to measure how well the attacked image represents the original prompt. We use the open-source model ViT-g-14~\cite{Radford2021OpenAIClip,ilharco_gabriel_2021_5143773} pretrained on LAION-5B~\cite{schuhmann2022laion5bopenlargescaledataset} to generate image and textual embeddings.  

\para{FID} measures the distance between probability distributions of images by mapping them to a latent representation from a predefined network InceptionV3~\cite{Szegedy_2016_Inception}, capturing similarities in the distributions rather than individual images.

\para{LPIPS}~\cite{zhang2018perceptual} computes the similarity of the activations generated by an image classification model at each layer, in our case Alexnet~\cite{NIPS2012_Alexnet}. Unlike FID, LPIPS is defined between two images rather than between two samples of images.

%\mjm{I commented out this aggregate metric below because it wasn't being used.}
In our evaluation, we are interested in the degradation effect of our attacks relative to a ground truth image (or set of images for FID). As such, for each of the image degradation metrics above, we calculate:
\begin{equation}
    | Q(x_0,x_{\text{ref}}) - Q(x^*_0,x_{\text{ref}}) |,
\end{equation}
where $Q$ is the quality metric (CLIP Score, FID, or LPIPS), $x_0$ is a generated watermarked image using \TR, $x^*_0$ is the transformed image after applying an attack, and $x_{\text{ref}}$ is the real image from ImageNet1k which was used to derive the prompt to generate $x_0$ and serves as the reference image.

\subsection{Evaluation Baselines}
We take various baselines as a reference point to assess the success of our attacks. These baselines include idealized settings and state-of-the-art attacks (\cite{an2024waves,lukas2024leveraging}).

\para{No-Attack:} \TR's performance with no attacks applied.

\para{Raw Pixel Values:} Directly train a surrogate detector on the images without converting to latent space, i.e. the setting considered in An et al.'s study~\cite{an2024waves}.

\para{True Latent Vectors:} Train a surrogate detector on the final ($t=0$) latent vectors before decoding the latents into images via AutoencoderKL. This is an idealized baseline as it assumes that the adversary has full access to the model instead of using the VAE to approximate these latents.

\para{Adversarial Noising:} We reproduce the adversarial noising attack by Lukas et al.~\cite{lukas2024leveraging}. This is also an idealized baseline as it assumes access to a model with the same architecture as the victim's, making it less practical than our attacks. We discard adversarial compression also by Lukas et al.\ because they show that it performs worse than adversarial noising~\cite{lukas2024leveraging}.

%\smallskip
%\noindent Next, we compare our attacks with these baselines. %We train the surrogate detector on the latent vectors after encoding the images via AutoencoderKL and applying FFT to transform them into the frequency domain.

\section{Evaluation}\label{sec:evaluation}
In this section, we present the results of our evaluation. We evaluate various aspects of the attacks, including their performance compared to the baselines, and the detector's precision under a range of deployment conditions. In addition, we conduct ablation studies to evaluate the impact of differences in the attack setup and the deployment setting, including a different VAE and a different diffusion model.

%\mjm{introduce section and summarize results. Our evaluation shows that the attacks...}

\subsection{No-Attack Baseline}
%\mjm{we can probably drop this section (at least the table, perhaps keep the discussion?}
For the first experiment, we test the \TR watermark detector on our dataset when no attack is in place. The results of this experiment are presented in the first row of \Cref{tab:adv_surr_attack}.

%\mjm{we should include a row in all the tables with this no-attack baseline so that it is easier for the reader to compare (instead of having a separate table).}
%\mjm{We should include all the metrics listed in the previous section in this table: PR-AUC, LPIPS, FID are missing.}
% \begin{table}[!ht]
%     \caption{Baseline Watermark Detection results}
%     \label{tab:baseline}
%     \centering
%     \begin{tabular}{cccc} \toprule
%          \textbf{ROC-AUC} & \textbf{Accuracy} & \textbf{TPR@1\%FPR} & \textbf{CLIP Score} \\ \midrule
%          1.0 & 1.0 & 1.0 & 0.364 \\ \bottomrule
%     \end{tabular}
% \end{table}

% As we can see, the performance of the \TR watermark detector achieves near perfect detection accuracy, with no impact on image quality degradation. We should expect there to be no impact on image quality degradation because no changes have been made to the images yet. These results serve as a baseline for assessing how the attacks impact the detector's performance in subsequent experiments.
The \TR watermark detector shows near perfect detection accuracy without affecting image quality. The high detection accuracy aligns with the original reported results, and no image degradation occurs since the images remain unchanged. This establishes a baseline for evaluating how attacks impact detector performance in later experiments.

% \begin{figure}[!ht]
%     \centering
%     \includegraphics[width=.8\columnwidth]{baseline_pcatbr.pdf}
%     \caption{Precison at different Base Rates for \TR watermarking.}
%     \label{fig:Pr@Br}
% \end{figure}

\subsection{Surrogate Detector Performance}
Before applying the attacks, we have evaluated the classification accuracy of the surrogate detectors we consider. We measure the training and validation accuracies to assess the potential overfitting of the surrogate detector. If the surrogate detector overfits the training data, the PGD attack is less likely to succeed in finding images that evade it. 

\begin{table}[t]
    \centering
    \caption{Surrogate Detector Training. Watermarked images (Wm) are generated using the victim's diffusion model with the \TR scheme. Non-watermarked images (UnWm) are generated without \TR. Public images (Pub) are taken from the ImageNet1k ILSVRC2012 dataset.\label{tab:surrogate_training}}
    \vspace{0.3cm}
    \resizebox{\linewidth}{!}{
    \begin{NiceTabular}{llcc}
    \CodeBefore
        \rowcolor{gray!15}{6-8}
    \Body
    \toprule
         \textbf{Attack Name} & \textbf{Training Dataset} & \makecell{{\textbf{Training}}\\{\textbf{Accuracy}}} & \makecell{{\textbf{Validation}}\\{\textbf{Accuracy}}} \\ \midrule
         \Block{2-1}{Raw Pixel Values} & Wm \& UnWm & 99.95\% & 96.44\% \\
         & Wm \& Pub & 99.86\% & 99.52\% \\ \midrule

         \Block{2-1}{True Latent Vectors} & Wm \& UnWm & 99.00\% & 96.00\% \\
         & Wm \& Pub & 99.93\% & 100\% \\ \midrule

         %\Block{2-1}{Guided Diffusion Images} & Wm \& UnWm & 98.12\% & 98.00\% \\
         %& Wm \& Pub & 87.65\% & 96.67\% \\ \midrule
        
        %\rowcolor{gray!15}
        
        \Block{2-1}{VAE-Recovered Latent Vectors} & Wm \& UnWm & 100\% & 94.78\% \\
         & Wm \& Pub & 99.93\% & 98.48\% \\ \bottomrule
    \CodeAfter
    \end{NiceTabular}}
\end{table}

As seen in \Cref{tab:surrogate_training}, training and validation accuracies are high for all surrogate detectors, with none of them significantly overfitting.
All surrogate detectors perform better when trained on Wm \& Pub than Wm \& UnWm, which can be explained by the greater difference between the watermarked and public image distributions in the latter dataset.
The detector trained on true latent vectors (second row) achieves the best validation accuracy, which is rather unsurprising given that there is no loss of information in approximating the latent.

The last row in \Cref{tab:surrogate_training} shows the results for our surrogate detector, namely the same task as before but training on latent vectors recovered via the AutoecoderKL. The drop in Validation Accuracy indicates that this task is more challenging. We hypothesize that this is due to the information loss incurred by re-encoding the image back into latent space with the VAE.

\subsection{VAE-Based Latent Vector Attacks}
After training all our surrogate models, we attempt a PGD attack on the watermarked images. \Cref{fig:adv_attack_egs} in Appendix~\ref{app:attack_egs} illustrates the visual impact of all attacks on an example image by providing the resulting attacked image (first column), and the difference between the original and resulting images in image space (second column) and Fourier space (third column). As shown in that figure, none of the attacks exhibits obvious unusual patterns in the image. However, the attacks have different impacts on the region of the Fourier space where the watermark is embedded, with our VAE-recovered attacks having a precise and concentrated effect on the key ring area.

\begin{table*}[t]
    \caption{Average performance of the attacks across three random seeds. Standard deviations were consistently below 0.01 for all metrics and are thus omitted. For all the scores, the lower the score, the more successful the attack is.\label{tab:adv_surr_attack}}
    \vspace{0.3cm}
    
    \resizebox{\linewidth}{!}{
        \begin{NiceTabular*}{1.135\linewidth}{@{} llccccccc @{}}
        \CodeBefore
            \rowcolor{gray!15}{12-13}
        \Body
        \toprule
             \bf{Attack Name} & \textbf{Training Data} & \textbf{PR-AUC} & \textbf{ROC-AUC} & \textbf{Accuracy} & \textbf{TPR@1\%FPR} & \textbf{CLIP Score} & \textbf{FID} & \textbf{LPIPS} \\
        \midrule
             \raisebox{0.1ex}{No Attack} & \raisebox{0.1ex}{N/A} & \raisebox{0.1ex}{0.994} & \raisebox{0.1ex}{0.993} & \raisebox{0.1ex}{0.965} & \raisebox{0.1ex}{0.968} & \raisebox{0.1ex}{0.000} & \raisebox{0.1ex}{0.000} & \raisebox{0.1ex}{0.000}\\\cline{1-9}
             \raisebox{-0.45ex}{Adversarial Noising} & \raisebox{-0.45ex}{N/A} & \raisebox{-0.45ex}{0.584} & \raisebox{-0.45ex}{0.459} & \raisebox{-0.45ex}{0.575} & \raisebox{-0.45ex}{0.141} & \raisebox{-0.45ex}{0.005} & \raisebox{-0.45ex}{0.602} & \raisebox{-0.45ex}{0.018}\\ \midrule
             \Block{2-1}{Raw Pixel Values} & Wm \& UnWm & 0.870 & 0.816 & 0.785 & 0.529 & 0.005 & 1.163 & 0.014 \\
             & Wm \& Pub & 0.828 & 0.743 & 0.756 & 0.485 & 0.006 & 0.303 & 0.008 \\ \midrule
             \Block{2-1}{True Latent Vectors} & Wm \& UnWm & 0.599 & 0.413 & 0.610 & 0.212 & 0.006 & 0.689 & 0.008 \\
             & Wm \& Pub & 0.742 & 0.615 & 0.692 & 0.343 & 0.005 & 0.519 & 0.004 \\ \midrule
             \Block{2-1}{SDXL-VAE Latent Vectors} & Wm \& Unwm & 0.452 & 0.333 & 0.516 & 0.039 & 0.023 & 33.71 & 0.143 \\
             & Wm \& Pub & 0.626 & 0.540 & 0.587 & 0.125 & 0.014 & 2.453 & 0.025 \\\midrule
             \Block{2-1}{16-Channel VAE Latent Vectors} & Wm \& Unwm & 0.834 & 0.767 & 0.752 & 0.450 & 0.008 & 3.754 & 0.021 \\
             & Wm \& Pub & 0.846 & 0.774 & 0.768 & 0.486 & 0.007 & 0.088 & 0.008 \\\midrule
             \Block{2-1}{VAE-Recovered Latent Vectors} & Wm \& UnWm & 0.350 & 0.108 & 0.508 & 0.023 & 0.009 & 2.684 & 0.021 \\
             & Wm \& Pub & 0.385 & 0.153 & 0.515 & 0.039 & 0.008 & 0.694 & 0.012 \\ \bottomrule
        \CodeAfter
        \end{NiceTabular*}%
    }
\end{table*}

In \Cref{tab:adv_surr_attack}, we report the performance of the watermark detector under each of the attacks, as well as the impact of such attacks on image quality.
We observe that our attack based on the recovery of the latents with AutoencoderKL is the most successful in disrupting watermark detection (highlighted row). Furthermore, the quality of the image is retained compared to the baseline CLIP score: it only rises by 0.008. Similarly, LPIPS and FID Score only have small increases, especially for the surrogate trained for distinguishing watermarked latents vs.\ public image latents. The changes in pixel space in the fifth and sixth rows of \Cref{fig:adv_attack_egs} are imperceptible.

Contrary to the expectation that access to the true latents (fourth row of \Cref{tab:adv_surr_attack}) would lead to better attack performance, we see that this attack performs worse than our surrogate attack trained on VAE-recovered latent vectors. We believe that  differences in distribution and scale between the true latent vectors and the VAE-recovered ones account for the attack's lower performance in this setting, as it learns an image perturbation that is effective in the true latent space but not necessarily in the VAE-recovered space.

As a result of the attacks, some of the ROC-AUC values are below the random guess baseline of $0.5$. %This happens because the attack causes the detector to classify watermarked images as non-watermarked.
Note that in an adversarial setting, the model owner cannot simply invert the detector to achieve ROC-AUC $ > 0.5$ because they have no reliable way of predicting the occurrence and the extent of an attack. Since the detector's decision is based on a threshold on the error between the recovered watermark and the initial watermark, flipping the decisions implies that the greater the dissimilarity between the recovered and initial watermark, the more likely it is to be watermarked. Therefore, if the model owner preemptively deploys the inverted detector and no attack takes place, inverting it causes it to miss most watermarked images, rendering the detector ineffective.
%the model owner cannot simply invert the predictions to improve performance over the random baseline, as the model owner cannot anticipate the adversary's strategy.
%Note that, in general, a classifier with ROC-AUC below the random guessing baseline can be converted into a classifier above the baseline by inverting its predictions.
%However, the model owner cannot anticipate the adversary's strategy.
%If the adversary targets only a few images and the model owner inverts the predictions, the detector would miss unattacked watermarked images, allowing adversaries to exploit them without the need for an attack.

\begin{figure}[!ht]
    \centering
    \includegraphics[width=1\columnwidth]{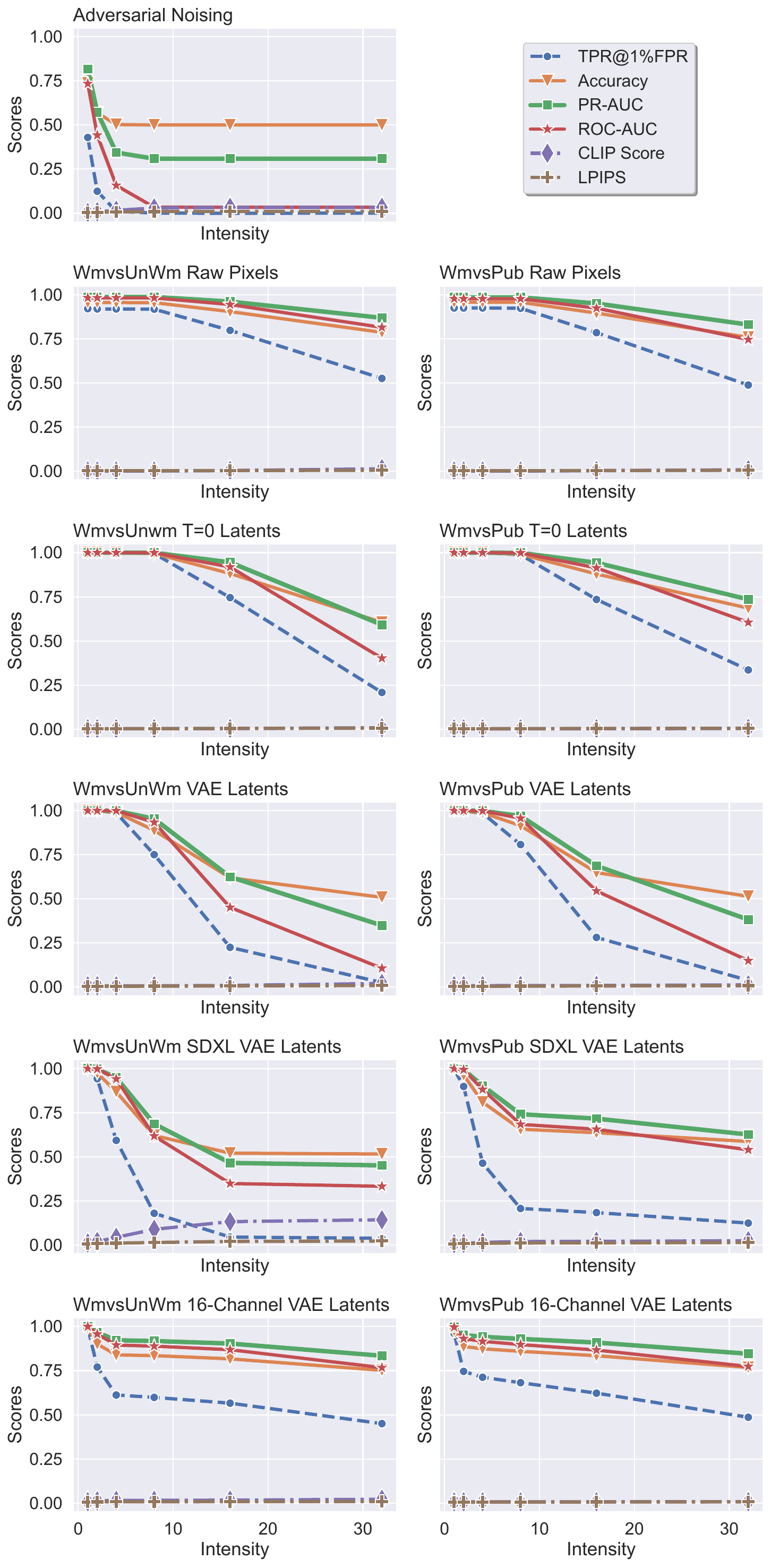}
    \vspace{-0.4cm}
    \caption{Results of our attack while varying $\delta$ to adjust the strength of the attack. Next to each score in the legend, we denote what direction of magnitude indicates a successful attack. We only include metrics that range within the unit interval, as such we exclude FID.}
    \label{fig:adv_attacks_strs}
    \vspace{-0.4cm}
\end{figure}

We found a discrepancy between An et al.'s and our results when executing the PGD attack with a surrogate trained on the Raw Pixel Values of the Wm \& Pub dataset: they reported an average TPR@1\%FPR of 0.99, but our results show a value of 0.48.
The main difference between our setups is that we trained the Resnet18 from the ground up while they fine-tuned a pretrained version. Using the pretrained Resnet18 may have caused their surrogates to incorporate broader features when fine-tuning for classification, which indicates that the adversary can benefit from training Resnet18 from scratch.

We reproduced the original adversarial noising paper's results, achieving a TPR@1\%FPR of 0.141 for $\delta=2$, slightly higher than the reported 0.052, likely due to dataset differences. Although adversarial noising minimally degrades image quality, our attacks are more effective and do not require access to the watermarking scheme.

We extend our attack evaluation by varying the perturbation budget, $\delta$, for each attack. These results are plotted in \Cref{fig:adv_attacks_strs}. The adversarial noising attack is extremely effective when using a $\delta$ of 8 and above. However, the image quality starts to deteriorate at that point. Our attacks work well with a $\delta$ of 32. Furthermore, image degradation is not affected for any of the surrogate models we have trained, demonstrating the effectiveness of surrogate detector attacks.

Finally, comparing these results with those reported in the original \TR paper, all the attacks included in our evaluation are significantly more effective than any of their tested approaches. They reported on average for each of their adversarial examples, an ROC-AUC of 0.975 and TPR@1\%FPR of 0.694. Compared with our reported results, we have an average of 0.519 ROC-AUC and 0.261 TPR@1\%FPR, which represents a 1.9x decrease in general detection accuracy. 

% \mjm{there are more figures on the table that should be discussed}

\subsection{Ablation Study for Different VAEs}\label{sect:abblation}
% \mjm{the results of this experiment should go here}
%then further show that the VAE chosen does not necessarily matter
While it is extremely common for text-to-image models to directly use a publicly available checkpoint, there still exist a few fine-tuned autoencoders that continue training from said publicly available checkpoint on Civitai. To account for this, we also evaluated our attack's effectiveness when using different VAEs to train our surrogate detectors, addressing scenarios where model owners might employ private VAEs. Specifically, we considered two cases: (1) when the victim model uses a fine-tuned version of a publicly available VAE, and (2) when the adversary uses a VAE with a different architecture. For the fine-tuned VAE scenario, we utilized SDXL's VAE, which is derived from Stable Diffusion's VAE. To test a different architecture, we employed a 16-Channel VAE~\cite{Ostris}, which has fewer parameters than Stable Diffusion's VAE and produces latents of dimension (64,64,16), compared to Stable Diffusion's (64,64,4) latent dimensions.

The results are presented in rows 5--6 of \Cref{tab:adv_surr_attack} and \Cref{fig:adv_attacks_strs}. Using SDXL's VAE, the attack remains robust, albeit less effective than when using the victim model's VAE. However, the impact on image quality becomes more pronounced. Training on the Wm \& UnWm dataset significantly degrades image quality, with the FID score increasing by 33.71 and similar increases in LPIPS and CLIP Score. We also show the attacked image in \Cref{fig:adv_attack_egs} of Appendix~\ref{app:attack_egs}. Upon close inspection, the image has noticeable image artifacts. The 16-Channel VAE's performance is comparable to training the surrogate detector on raw images. Interestingly, when trained on the Wm vs. Pub dataset, it causes less image degradation. We hypothesize that SDXL's superior performance over the 16-Channel VAE is due to it being a fine-tuned version of the victim model's VAE, resulting in re-estimated latents with distributions similar to those of the victim model's VAE.

These findings suggest that using a customized VAE significantly enhances watermark robustness. We observed that increasing the perturbation budget $\delta$, leads SDXL VAE to substantial image degradation, particularly for the surrogate trained on the Wm \& UnWm dataset. We anticipate that further increasing $\delta$ would eventually impact the quality of attacked images derived from the surrogate trained on the Wm \& Pub dataset as well. This implies a trade-off between attack effectiveness and maintaining image quality, with the choice of VAE playing a crucial role in this balance.

\subsection{Ablation Study Without a VAE}~\label{sect:abblation_model}
To determine the importance of having access to a VAE for the success of our attack, we evaluate it on OpenAI's Guided Diffusion~\cite{dhariwal}, an open source implementation of a diffusion model that does not use a VAE. While \TR can be directly applied to Guided Diffusion, this is not a text-to-image model but a class-conditioned diffusion model trained on ImageNet classes and thus does not expect a prompt for generation. Aside from this, the evaluation follows the same methodology: we use the model to generate images and construct the Wm \& Unwm and Wm \& Pub datasets to train the surrogate models and apply PGD on them.
%Guided Diffusion is also used as a baseline in \TR thus we can be sure that this model is suitable for Tree Ring Watermarking.

\begin{table}[h]
    \centering
    \caption{Attack performance on \TR applied to Guided Diffusion. Unlike \Cref{tab:adv_surr_attack}, we omit CLIP Score and LPIPS as Guided Diffusion is a class-conditioned model and ground truth images are not available for assessing generation quality.\label{tab:guided_diffusion}}
    \vspace{0.3cm}
    \resizebox{\linewidth}{!}{
    \begin{NiceTabular}{lccccc}
    \CodeBefore
    \Body
    \toprule
         \textbf{Training Data} & \textbf{PR-AUC} & \textbf{ROC-AUC} & \textbf{Accuracy} & \textbf{TPR@1\%FPR} & \textbf{FID} \\ \midrule
         Wm \& UnWm & 0.8503 & 0.7870 & 0.7713 & 0.4880 & 56.44 \\ \midrule
         Wm \& Pub & 0.8506 & 0.7855 & 0.7712 & 0.4803 & 56.44 \\ \bottomrule
    \CodeAfter
    \end{NiceTabular}}
\end{table}

\Cref{tab:guided_diffusion} presents the results of this evaluation. While the attack still decreases the effectiveness of \TR, it is not as successful as when the VAE is available. This result aligns with the findings by An et al.~\cite{an2024waves}, who show that surrogate-based attacks on images do not perform as well as other adversarial attacks, indicating that the attack achieves lower performance if the attacker does not use a VAE to encode the images into the latent space. These results highlight the critical role of the VAE in enabling our attack and demonstrate the risk posed by making this diffusion model component publicly available.

%We can see that training on the Wm \& Pub dataset is less definitive as with the other data types. The difficulty in reliably classifying an image as watermarked or not will heavily impact our ability to execute the attack as it suggests that the surrogate did not learn characteristics of the watermark.

%PGD is heavily dependent on the accuracy of the surrogate model. If the surrogate model misclassifies a watermarked image as non-watermarked, the projected gradients are vanishingly small due to a near-zero error between the predicted and target classes. This explains the reduced attack performance on the Wm \& Pub. However, the accuracy of the surrogate model trained on Wm \& UnWm is almost on par with that of the other surrogate models. 

\subsection{Impact of the Base Rate}\label{sec:base_rate}
Precision has been overlooked in the evaluation of watermarking schemes. This is a limitation, as it misses an aspect of detection performance not captured by performance metrics based on TPRs and FPRs. Even with high TPR and low FPR, precision can be low if the base rate of the positive class (watermarked) is low, leading to low confidence in the detection of watermarked images. In real-world deployments, the base rate is likely to be low, as any watermarking scheme that is not broadly adopted will be overwhelmingly exposed to non-watermarked content.
Excluding precision in the evaluation of a detector where a class imbalance is likely is a case of the \emph{base rate fallacy}, an evaluation bias that has been extensively studied in domains with similar detection tasks~\cite{axelsson1999base,juarez2014critical,arp2022and}.

% Finally, we evaluate the impact of the \emph{base rate} on the \TR detector's precision. The base rate is the prior probability of a watermarked image, which is likely to be low in a real-world deployment of a watermarking scheme; at least initially, the watermark detector will be exposed overwhelmingly to non-watermarked content.
% Not including precision in the evaluation can be critical under a low base rate setting because precision is a function of it while other metrics like accuracy and TPR@1\%FPR are not, missing a crucial aspect of the detector's performance and reliability.
% In the literature on ML watermarking, precision-based metrics have been overlooked, with most studies using a balanced dataset in their evaluation, and thus implicitly assuming a uniform base rate.

\begin{figure}[!ht]
    \centering
    \includegraphics[width=1\columnwidth]{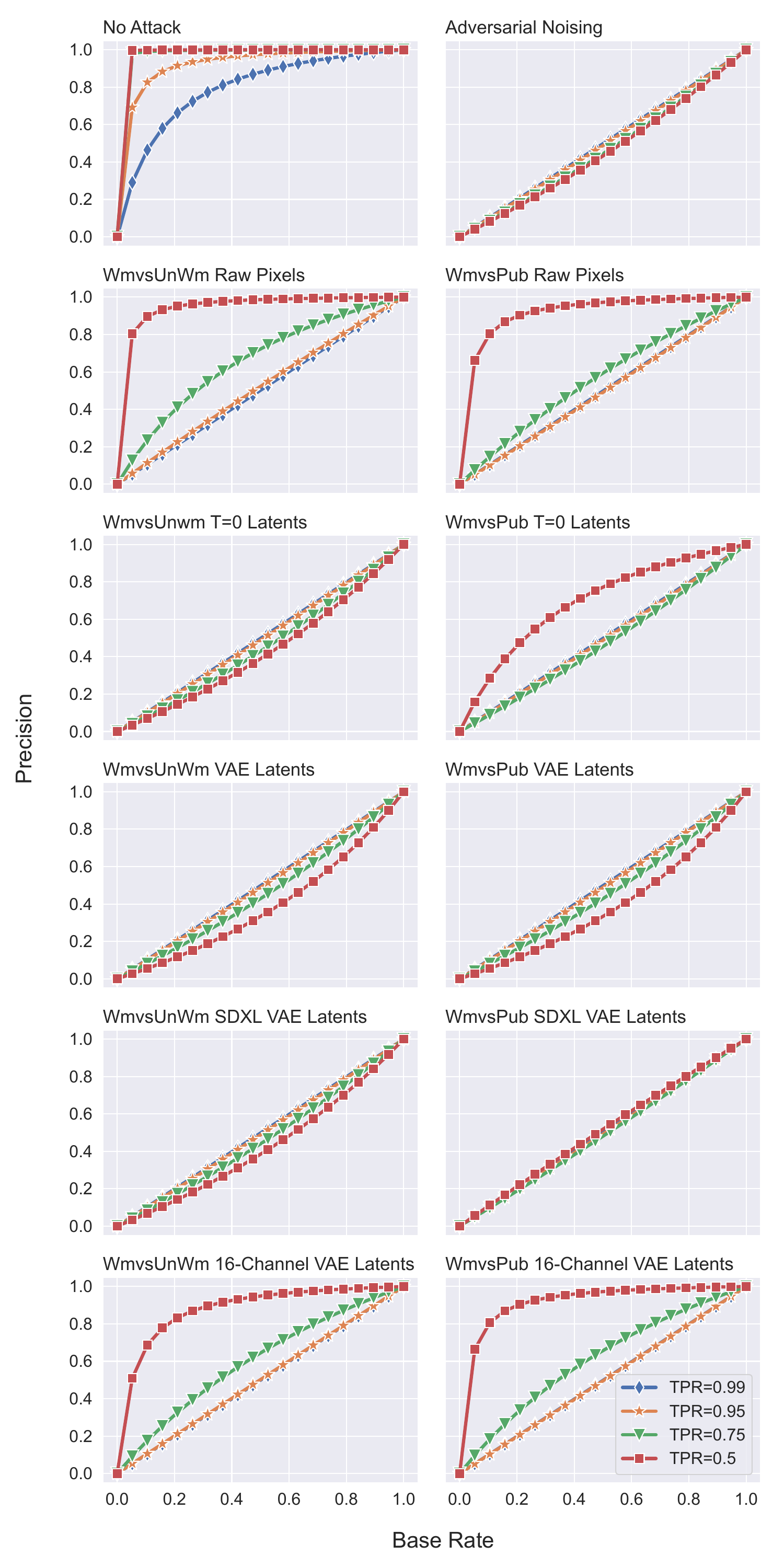}
    \vspace{-0.6cm}
    \caption{Detector precision over the base rate for each attack. Each line represents a different operating point in the ROC. \label{fig:Pr@Br_all}}
   
    \vspace{-0.4cm}
\end{figure}

%~\cite{axelsson1999base}.
Since we do not know the base rate of real-world deployments, we plot the precision curves of all attacks and baselines over a range of possible base rates for four ROC operating points obtained by imposing a minimum TPR of 0.5, 0.75, 0.95, and 0.99 (see \Cref{fig:Pr@Br_all}). 
The linearity of the high-TPR curves is expected, as in those cases TPR$\approx$FPR.
However, for base rates below 0.1, which are plausible in a deployment scenario, precision decreases significantly, even for high TPR.

In the fourth row of \Cref{fig:Pr@Br_all} we see the precision curves resulting from applying our attack. For both datasets, the false positives induced by our attacks make precision drop linearly even for TPR=0.5.
\Cref{fig:prc} of Appendix~\ref{app:pr_curves} complements these results with the Precision--Recall curves where we can see that our attacks are particularly effective in decreasing the precision of the \TR detector. 

Given these results, we would not recommend this scheme for deployment. Even in the non-adversarial setting, for a base rate of 0.1, the model owner would require picking an operating point close to TPR = 0.5 to achieve high precision (see also the first column of \Cref{fig:prc} in Appendix~\ref{app:pr_curves}), missing half of the total watermarked images.
%Portraying \TR watermarking as effective without measuring its precision under realistic base rates is a case of the base rate fallacy. This bias in previous evaluations may have overestimated the detector's performance in practice.
\section{Discussion}
We now discuss the implications of our findings, not only for \TR, but also for other image watermarking schemes.

\para{Attack's reliance on VAE access.}
Our results indicate that the attack heavily relies on the target VAE. We conducted two ablation studies to test this assumption: one where the attacker uses a different VAE (\Cref{sect:abblation}), and another where the diffusion model does not use a VAE at all (\Cref{sect:abblation_model}). In both settings, the attack is significantly less effective compared to the ideal case. Despite this limitation, our contributions remain practically relevant, as these ideal conditions are likely to be met in real-world scenarios.
As argued in \Cref{sect:threat_model}, it is common practice to reuse existing VAEs to train diffusion models through latent diffusion.

\para{The security arms race.}
As watermarking schemes continue to gain traction in both academia and industry~\cite{synthid2024}, ensuring their robustness against removal becomes critical. Currently, most security claims for these schemes are based on experimental evaluations against ad-hoc attacks, meaning they are only considered secure until a successful attack is discovered. The security of image watermarking schemes should instead be grounded in computational hardness assumptions or existing cryptographic primitives.
There is progress in this direction for watermarking schemes for large language models~\cite{christ2024,liu2023,fairoze2023} and, more recently, for image models~\cite{gunn2024undetectable}. 

\para{Precision as a key metric.}
The evaluation of watermarking schemes often focuses on metrics that do not include precision. However, relying solely on these metrics may overestimate the performance of the detector once it is deployed. Thus, security evaluations of image watermarking should use precision-based metrics like PR-AUC, reflecting performance in low base rate scenarios, such as those expected in the future deployment of these watermarking schemes.

Our evaluation of \TR's precision reveals that, even in the absence of attacks, the detector's precision is low in realistic deployment scenarios. In a world where most images do not contain a watermark, the detector's confidence in detecting watermarks decreases significantly without an attack and falls drastically under our attacks, indicating that the detector is not reliable when it detects watermarks in general settings.

These findings suggest that if watermarking schemes like \TR were deployed to monitor content on publicly accessible platforms, as it is currently proposed, their performance would be inadequate at best and completely unusable if attacks were widespread. This raises concerns about the practical viability of these watermarking techniques.

%Despite its recognition in the security community, we find that this problem is largely overlooked in the literature on ML watermarking.

%\para{Low precision under realistic conditions.}

\para{Potential countermeasures.}
Our attacks exploit the shift in the distribution of the latent space caused by the embedding of a fixed watermark. Therefore, a natural mitigation strategy is to ensure that the latent representations of watermarked and non-watermarked images are indistinguishable, making it harder to train effective surrogate detectors.

Additionally, the success of our attacks degrades when the attacker and the victim’s autoencoder differ. Thus, a short-term countermeasure is to train diffusion models with a custom private autoencoder, ensuring some degree of misalignment between the diffusion latents and those estimated by the attack, although this would mean that practitioners would not be able to benefit from the cost savings offered by latent diffusion~\cite{an2024waves}.

Defenders may also leverage asymmetries in the deployment setting. For example, \TR assumes the attacker has black-box access to the diffusion model, as knowledge of its weights would enable direct recovery of the watermarks. 
This restricted access forces the adversary to rely on techniques that approximate the inversion process, which increases the cost of attack compared to that of defense. To capitalize on this asymmetry, a model owner could embed multiple watermarks per image---potentially using different embedding schemes and rotating the keys---thereby forcing an attacker to craft perturbations that evade several detectors simultaneously.
In addition, access-control mechanisms could restrict the number of queries associated with a specific watermarking key, thus limiting the adversary's ability to collect a sufficient number of watermarked samples for training a surrogate detector.

Countermeasures designed against adversarial examples, such as adversarially training the watermark detector, would improve the robustness of the detector against surrogate PGD attacks. However, prior work has consistently shown that such defenses often fail against adaptive attackers~\cite{pmlr-v80-athalye18a} or significantly degrade model accuracy~\cite{cohen2019certified}.
Currently, no defense provides certified robustness to adversarial examples without incurring substantial accuracy trade-offs.

%Although model stealing attacks would undermine the security of \TR and other in-processing techniques, they are also computationally costly and can be 

% \jun{I would probably just say with custom private ones here. I haven't tried it yet but if \Cref{fig:w_w/o_latent} is anything to go by, If we generate images without a VAE, its very likely if someone does FFT on an image, the remnants of the watermark will be publicly visible.}
% \mjm{good point! removed it.}

\para{External validity of the results.}
Note that we do not require evaluating the attacks on multiple datasets to demonstrate the generality of the attacks, as the attacker can choose any publicly available dataset to train the surrogate detector. %By using ImageNet, we are thus assuming an attacker that follows the most convenient method possible.

Although we have not tested the effectiveness of our attacks against other watermarking schemes, our analysis indicates that any watermarking scheme that embeds a distinctive watermark in the latent space is likely to be susceptible to a VAE-based attack, even if the difference is not visible in the image space.

%By using ImageNet, we are thus assuming a scenario in which the attacker is building a dataset via the most convenient method possible. %A popular image dataset like ImageNet does not include images watermarked with \TR which is suitable to model the negative class.

\section{Conclusion}
Our findings have uncovered a novel class of surrogate detector attacks based on public pretrained autoencoders. The results demonstrate that our method not only surpasses existing surrogate detector attacks but also outperforms other attacks that make strong assumptions about the threat model. Additionally, our evaluation shows that \TR's performance is lower than initially thought under typical deployment scenarios, raising concerns about the practicality of these schemes.

\section*{Acknowledgments}
We thank our shepherd and the anonymous reviewers for their valuable feedback.
This research was supported by the LFCS internship program at the University of Edinburgh, which funded Junhua Lin, and by the University of Edinburgh's Generative AI Laboratory (GAIL), which provided access to their high-performance cluster. Marc Juarez is a GAIL fellow and the recipient of a Google Research Scholar Award in Security.
We are grateful to our colleagues who provided feedback on earlier drafts of the manuscript, including Kai Yao and Rik Sarkar. We extend special thanks to Nils Lukas for his assistance in reproducing adversarial noising, and to Kai Yao and Bence Szilágyi for their help in setting up the GAIL cluster and providing helper scripts.
%We are also grateful to our colleagues who offered feedback on earlier drafts of the manuscript, including Kai Yao and Rik Sarkar. Special thanks to Kai Yao, as well as Bence Szilágyi and Matyas Zsoldos from Garandor, for their assistance in setting up the GAIL cluster and providing helper scripts.
%\acks: Reviewers, LFCS internship program, GAIL (Luna), GAIL seed fund project, Google Research Scholar Award, Rik, Kai, Garandor (Bence and Matyas), Cryptosec group.

\section*{Ethical Considerations}
We have reviewed the USENIX Security Ethics Guidelines and carefully considered the ethical aspects of this research.

Regarding the impact of our findings, we concluded that the benefits of exposing these attacks clearly outweigh the risks. ML watermarking schemes are still in the prototype stage and have not been deployed. Therefore, the likelihood of malicious actors exploiting these attacks is minimal, while exposing their vulnerabilities gives a better understanding of the limitations of ML watermarking. 
It is crucial to raise awareness of the vulnerabilities of these schemes so that contingency measures are in place before they are deployed more broadly.
Furthermore, since attacks were evaluated in a controlled and isolated environment, conducting this research did not cause any harm.

%In addition, the conducted research adhered to standard practices in security research: All attacks were evaluated in a controlled and isolated environment, and, therefore, the research did not cause any harm. 

\section*{Open Science}
% In accordance with best practices for open science, we will make all artifacts produced by this work publicly available under open licenses so that future researchers can reproduce our findings. We will upload the source code for the experiments, including the attack implementations and data processing helper methods, to a public code repository. This repository will include detailed instructions for running the experiments and reproducing the results. The datasets we used, as well as the implementations of \TR and Stable Diffusion were already made publicly available by other researchers; we will provide instructions on how to download them and integrate them into our evaluation framework.
% The repository containing all artifacts will be made available for artifact evaluation upon acceptance of the paper.
In accordance with best practices for open science, we make our \href{https://doi.org/10.5281/zenodo.15595719}{source code}\footnote{\url{https://doi.org/10.5281/zenodo.15595719}} openly available. The repository contains our attack implementations, data processing helper methods, and the datasets used. All diffusion models used and \TR were made publicly available by other researchers; we have provided instructions on how to download and integrate them into our evaluation framework. We also provide instructions on how to reproduce the results of our experiments.

{\footnotesize \bibliographystyle{acm}
\bibliography{usenix}}

\input{appendix}
\end{document}

%% file: appendix.tex
\section*{Appendix}
\addcontentsline{toc}{section}{Appendices}
\renewcommand{\thesubsection}{\Alph{subsection}}
\subsection{Attack Visualization}\label{app:attack_egs}

\begin{figure*}[t]
    \centering
    \subfloat{
        \centering
        \includegraphics[width=.95\columnwidth]{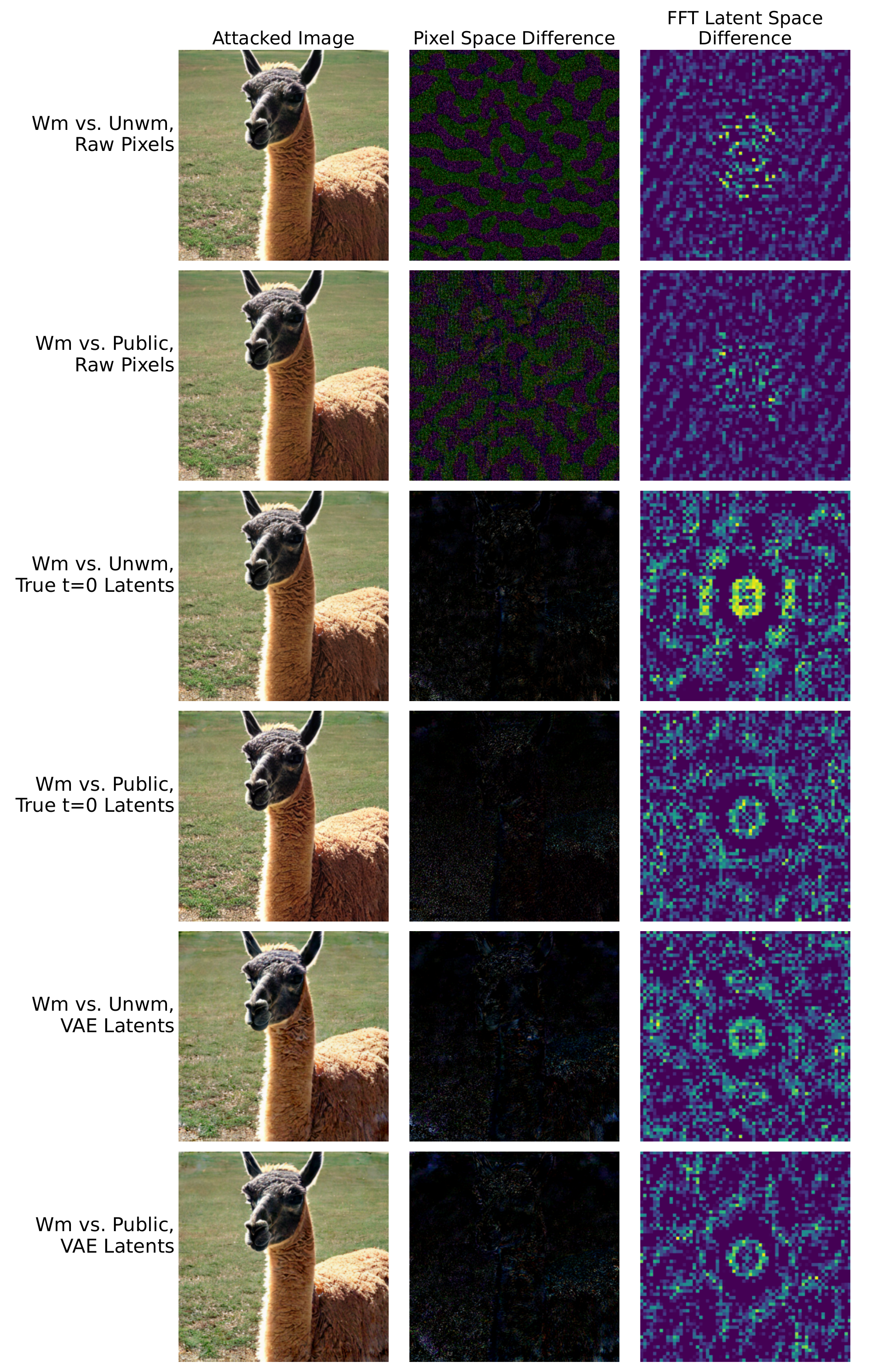}
    }
    \hfill
    \subfloat{
        \centering
        \includegraphics[width=.95\columnwidth]{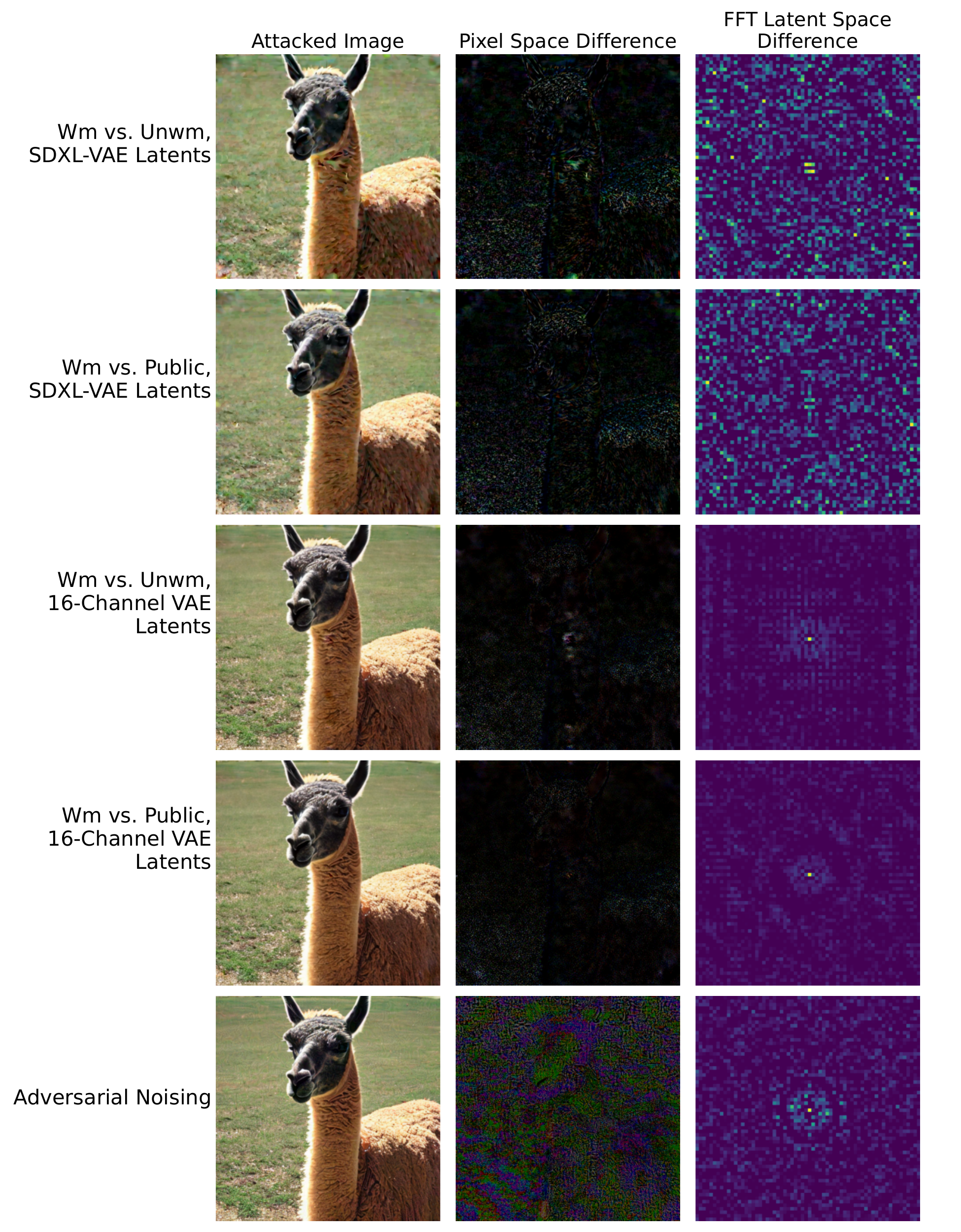}
    }
    \caption{Attacked images. Each row corresponds to a combination of dataset and input type. The columns from left to right are: our augmented images $x_{0_{aug}}$, the changes in pixel values, and the changes in latent space for the watermark channel after FFT. }
    \label{fig:adv_attack_egs}
\end{figure*}

From \Cref{fig:adv_attack_egs}, we can see that each attack learns a correlation between the area where the watermark is embedded and the class prediction. As seen in the difference between the Fourier transform of the attacked images and the original images (third column), the surrogates trained on the Wm \& UnWm dataset learn a significantly stronger correlation, which is unexpected given that the difference between the two latent distributions is only the watermark. With surrogate detectors trained on the Wm \& Pub dataset, the correlation is weaker, but compared to surrogates trained on raw pixel values, the attack is more concentrated around the key ring, rather than scattered throughout the space. When training on the true latents (at $t=0$), the correlation between attack area and misclassification becomes stronger; however, it also starts to include features outside the key ring.

When using a fine-tuned version of the victim's VAE, the SDXL-VAE, we see a more sparse effect of the attack in Fourier space, which is in line with the degradation in performance of the attack. Furthermore, when the victim uses a different VAE, the 16-Channel VAE, there is limited impact of the attack on the Fourier space. 
%\mjm{comment on VAE results of figure is missing}

\subsection{Precision--Recall Curves}\label{app:pr_curves}

\begin{figure*}[b!]
    \centering
    \includegraphics[width=\textwidth]{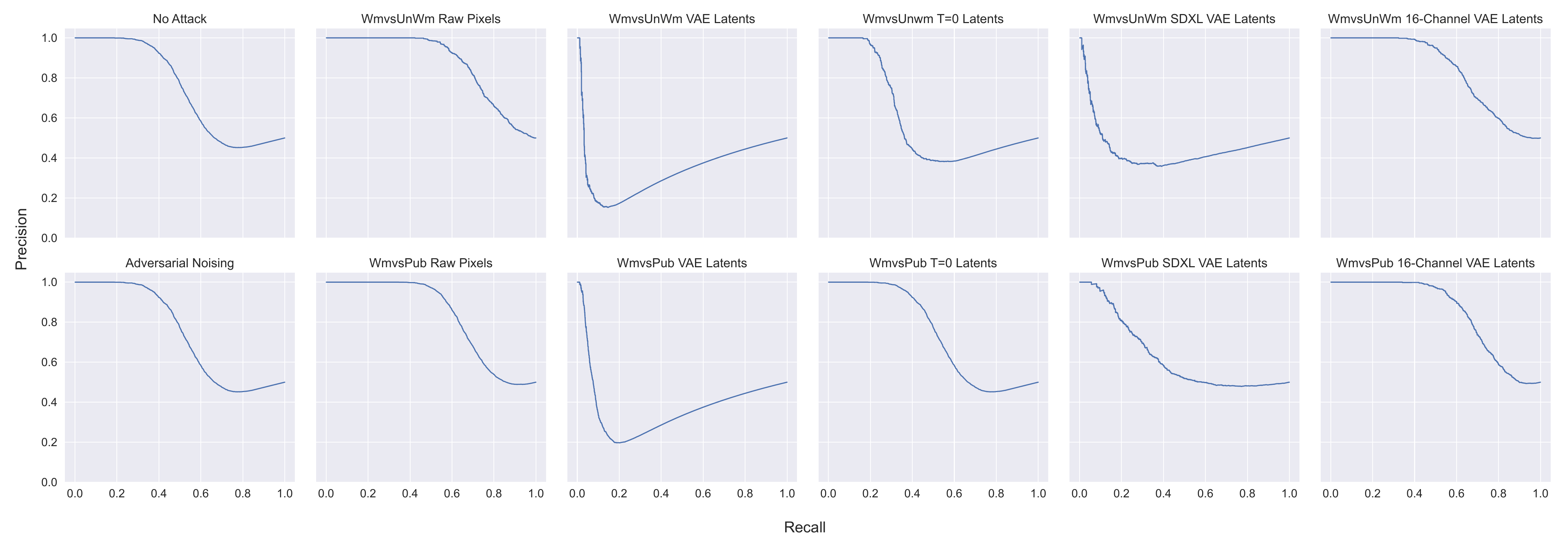}
    \vspace{-0.8cm}
    \caption{PR curves for all the attacks when the base rate is 0.5. $y=0.5$ is the PR curve of random guessing.}
    \label{fig:prc}
\end{figure*}

\Cref{fig:prc} plots the Precision--Recall curves for all the attacks at a base rate of $0.5$, namely a balanced dataset. Remarkably, our attack (third column) exerts the most significant impact on the curve, making it the attack in which the defender would achieve the weakest compromise between precision and recall. This is especially relevant in scenarios where the base rate is low, as the additional false positives can have an overwhelming effect on the deployment performance of the detector.\balance

% \jun{Can't use subcaption package in here}
% \begin{figure}
%     \centering
%     \includegraphics[width=\columnwidth]{second_img_grid.pdf}
%     \caption{Attacks visualized, each row of images corresponds to a combination of dataset and input type. First column shows our augmented images $x_{0_{aug}}$, the second column shows the changes in the pixel values, the third column shows the changes in latent space for the watermark channel after applying FFT. }
%     \label{fig:adv_attack_egs_2}
% \end{figure}

%% file: usenix.bbl
\begin{thebibliography}{10}

\bibitem{deepdreamgenerator}
{\sc {Aifnet}}.
\newblock {Deep Dream Generator}.
\newblock \url{https://deepdreamgenerator.com}.
\newblock Accessed: May 27, 2025.

\bibitem{aiseoart}
{\sc {AISEO}}.
\newblock {AISEO Art - AI art generation}.
\newblock \url{https://art.aiseo.ai}.
\newblock Accessed: May 27, 2025.

\bibitem{al2007combined}
{\sc Al-Haj, A.}
\newblock {Combined DWT-DCT digital image watermarking}.
\newblock {\em Journal of computer science 3}, 9 (2007), 740--746.

\bibitem{an2024waves}
{\sc An, B., Ding, M., Rabbani, T., Agrawal, A., Xu, Y., Deng, C., Zhu, S., Mohamed, A., Wen, Y., Goldstein, T., and Huang, F.}
\newblock {WAVES: Benchmarking the Robustness of Image Watermarks}.
\newblock In {\em {International Conference on Machine Learning (ICML)}\/} (2024).

\bibitem{arp2022and}
{\sc Arp, D., Quiring, E., Pendlebury, F., Warnecke, A., Pierazzi, F., Wressnegger, C., Cavallaro, L., and Rieck, K.}
\newblock Dos and don'ts of machine learning in computer security.
\newblock In {\em 31st USENIX Security Symposium (USENIX Security 22)\/} (2022), pp.~3971--3988.

\bibitem{pmlr-v80-athalye18a}
{\sc Athalye, A., Carlini, N., and Wagner, D.}
\newblock Obfuscated gradients give a false sense of security: Circumventing defenses to adversarial examples.
\newblock In {\em Proceedings of the 35th International Conference on Machine Learning\/} (10--15 Jul 2018), J.~Dy and A.~Krause, Eds., vol.~80 of {\em Proceedings of Machine Learning Research}, PMLR, pp.~274--283.

\bibitem{axelsson1999base}
{\sc Axelsson, S.}
\newblock The base-rate fallacy and its implications for the difficulty of intrusion detection.
\newblock In {\em Proceedings of the 6th ACM Conference on Computer and Communications Security\/} (1999), pp.~1--7.

\bibitem{cogview4}
{\sc {BigModel}}.
\newblock {CogView 4 - Large Vision Language Model}.
\newblock \url{https://open.bigmodel.cn/pricing}.
\newblock Accessed: May 27, 2025.

\bibitem{bird2023typology}
{\sc Bird, C., Ungless, E., and Kasirzadeh, A.}
\newblock Typology of risks of generative text-to-image models.
\newblock In {\em Proceedings of the 2023 AAAI/ACM Conference on AI, Ethics, and Society\/} (2023), pp.~396--410.

\bibitem{flux}
{\sc {Black Forest Labs}}.
\newblock {FLUX 1.1 Pro}.
\newblock \url{https://huggingface.co/black-forest-labs/FLUX.1-dev}, 2024.
\newblock Accessed: May 27, 2025.

\bibitem{boenisch2021systematic}
{\sc Boenisch, F.}
\newblock A systematic review on model watermarking for neural networks.
\newblock {\em Frontiers in big Data 4\/} (2021), 729663.

\bibitem{Ostris}
{\sc Burkett, J.}
\newblock {Ostris} vae - kl-f8-d16, 7 2024.

\bibitem{christ2024}
{\sc Christ, M., Gunn, S., and Zamir, O.}
\newblock Undetectable watermarks for language models.
\newblock In {\em The Thirty Seventh Annual Conference on Learning Theory\/} (2024), PMLR, pp.~1125--1139.

\bibitem{clipdrop}
{\sc {Clipdrop}}.
\newblock {Clipdrop - AI tools powered by Stable Diffusion}.
\newblock \url{https://clipdrop.co}.
\newblock Accessed: May 27, 2025.

\bibitem{cohen2019certified}
{\sc Cohen, J., Rosenfeld, E., and Kolter, Z.}
\newblock Certified adversarial robustness via randomized smoothing.
\newblock In {\em international conference on machine learning\/} (2019), PMLR, pp.~1310--1320.

\bibitem{cox2007digital}
{\sc Cox, I., Miller, M., Bloom, J., Fridrich, J., and Kalker, T.}
\newblock {\em {Digital watermarking and steganography}}.
\newblock Morgan kaufmann, 2007.

\bibitem{craiyon}
{\sc {Craiyon}}.
\newblock {Craiyon - AI model based on DALL-E}.
\newblock \url{https://www.craiyon.com}.
\newblock Accessed: May 27, 2025.

\bibitem{davinci_sdxl}
{\sc {Davinci AI}}.
\newblock {Davinci AI - Powered by SDXL}.
\newblock \url{https://davinci.ai}.
\newblock Accessed: May 27, 2025.

\bibitem{synthid2024}
{\sc DeepMind, G.}
\newblock Synthid: Identifying ai-generated images.
\newblock \url{https://deepmind.google/technologies/synthid}, 2023.
\newblock Accessed: 2024-09-08.

\bibitem{dhariwal}
{\sc Dhariwal, P., and Nichol, A.}
\newblock Diffusion models beat gans on image synthesis.
\newblock In {\em Advances in Neural Information Processing Systems\/} (2021), M.~Ranzato, A.~Beygelzimer, Y.~Dauphin, P.~Liang, and J.~W. Vaughan, Eds., vol.~34, Curran Associates, Inc., pp.~8780--8794.

\bibitem{du2024sok}
{\sc Du, L., Zhou, X., Chen, M., Zhang, C., Su, Z., Cheng, P., Chen, J., and Zhang, Z.}
\newblock Sok: Dataset copyright auditing in machine learning systems.
\newblock {\em arXiv preprint arXiv:2410.16618\/} (2024).

\bibitem{fairoze2023}
{\sc Fairoze, J., Garg, S., Jha, S., Mahloujifar, S., Mahmoody, M., and Wang, M.}
\newblock Publicly-{{Detectable Watermarking}} for {{Language Models}}, 2023.

\bibitem{fernandez2023stable}
{\sc Fernandez, P., Couairon, G., J{\'e}gou, H., Douze, M., and Furon, T.}
\newblock The stable signature: Rooting watermarks in latent diffusion models.
\newblock In {\em Proceedings of the IEEE/CVF International Conference on Computer Vision\/} (2023), pp.~22466--22477.

\bibitem{gunn2024undetectable}
{\sc Gunn, S., Zhao, X., and Song, D.}
\newblock An undetectable watermark for generative image models.
\newblock {\em arXiv preprint arXiv:2410.07369\/} (2024).

\bibitem{he2015deepresiduallearningimage}
{\sc He, K., Zhang, X., Ren, S., and Sun, J.}
\newblock Deep residual learning for image recognition.
\newblock In {\em Proceedings of the IEEE Conference on Computer Vision and Pattern Recognition (CVPR)\/} (6 2016).

\bibitem{DDPM}
{\sc Ho, J., Jain, A., and Abbeel, P.}
\newblock Denoising diffusion probabilistic models.
\newblock In {\em Advances in Neural Information Processing Systems\/} (2020), H.~Larochelle, M.~Ranzato, R.~Hadsell, M.~Balcan, and H.~Lin, Eds., vol.~33, Curran Associates, Inc., pp.~6840--6851.

\bibitem{hu2021loralowrankadaptationlarge}
{\sc Hu, E.~J., Shen, Y., Wallis, P., Allen-Zhu, Z., Li, Y., Wang, S., Wang, L., and Chen, W.}
\newblock Lo{RA}: Low-rank adaptation of large language models.
\newblock In {\em The {{Tenth International Conference}} on {{Learning Representations}}\/} (2022).

\bibitem{huggingface}
{\sc Hugging~Face, I.}
\newblock {Hugging Face}.
\newblock \url{https://huggingface.co/}, 2016.
\newblock Accessed: May 21, 2025.

\bibitem{ilharco_gabriel_2021_5143773}
{\sc Ilharco, G., Wortsman, M., Wightman, R., Gordon, C., Carlini, N., Taori, R., Dave, A., Shankar, V., Namkoong, H., Miller, J., Hajishirzi, H., Farhadi, A., and Schmidt, L.}
\newblock Openclip, July 2021.
\newblock If you use this software, please cite it as below.

\bibitem{juarez2014critical}
{\sc Juarez, M., Afroz, S., Acar, G., Diaz, C., and Greenstadt, R.}
\newblock A critical evaluation of website fingerprinting attacks.
\newblock In {\em Proceedings of the 2014 ACM SIGSAC conference on computer and communications security\/} (2014), pp.~263--274.

\bibitem{kamali2024distinguish}
{\sc Kamali, N., Nakamura, K., Chatzimparmpas, A., Hullman, J., and Groh, M.}
\newblock How to distinguish ai-generated images from authentic photographs, 2024.

\bibitem{kingma2022autoencodingvariationalbayes}
{\sc Kingma, D.~P., and Welling, M.}
\newblock Auto-encoding variational bayes.
\newblock In {\em 2nd International Conference on Learning Representations, {ICLR} 2014, Banff, AB, Canada, April 14-16, 2014, Conference Track Proceedings\/} (2014), Y.~Bengio and Y.~LeCun, Eds.

\bibitem{NIPS2012_Alexnet}
{\sc Krizhevsky, A., Sutskever, I., and Hinton, G.~E.}
\newblock Imagenet classification with deep convolutional neural networks.
\newblock In {\em Advances in Neural Information Processing Systems\/} (2012), F.~Pereira, C.~Burges, L.~Bottou, and K.~Weinberger, Eds., vol.~25, Curran Associates, Inc.

\bibitem{leonardoai}
{\sc {Leonardo AI}}.
\newblock {Leonardo.Ai - Generative AI with Stable Diffusion}.
\newblock \url{https://leonardo.ai}.
\newblock Accessed: May 27, 2025.

\bibitem{liu2023}
{\sc Liu, A., Pan, L., Hu, X., Li, S., Wen, L., King, I., and Philip, S.~Y.}
\newblock An unforgeable publicly verifiable watermark for large language models.
\newblock In {\em The {{Twelfth International Conference}} on {{Learning Representations}}\/} (2023).

\bibitem{lu2023seeing}
{\sc Lu, Z., Huang, D., Bai, L., Qu, J., Wu, C., Liu, X., and Ouyang, W.}
\newblock Seeing is not always believing: benchmarking human and model perception of ai-generated images.
\newblock {\em Advances in Neural Information Processing Systems (NeurIPS) 36\/} (2024).

\bibitem{lukas2024leveraging}
{\sc Lukas, N., Diaa, A., Fenaux, L., and Kerschbaum, F.}
\newblock Leveraging optimization for adaptive attacks on image watermarks.
\newblock In {\em The Twelfth International Conference on Learning Representations, {ICLR} 2024, Vienna, Austria, May 7-11, 2024\/} (2024).

\bibitem{lukas2023ptw}
{\sc Lukas, N., and Kerschbaum, F.}
\newblock $\{$PTW$\}$: Pivotal tuning watermarking for $\{$Pre-Trained$\}$ image generators.
\newblock In {\em 32nd USENIX Security Symposium (USENIX Security 23)\/} (2023), pp.~2241--2258.

\bibitem{civitai}
{\sc Maier, J.}
\newblock {CivitAI}.
\newblock \url{https://civitai.com/}, 2022.
\newblock Accessed: Jan 21, 2025.

\bibitem{dalle-vae}
{\sc {OpenAI}}.
\newblock {openai/DALL-E}.
\newblock \url{https://github.com/openai/DALL-E}, 2021.
\newblock Accessed: May 27, 2025.

\bibitem{openai_sora}
{\sc {OpenAI}}.
\newblock {Sora: Text-to-Video Generation}.
\newblock \url{https://openai.com/sora}, 2024.
\newblock Accessed: May 27, 2025.

\bibitem{podell2023sdxl}
{\sc Podell, D., English, Z., Lacey, K., Blattmann, A., Dockhorn, T., M{\"{u}}ller, J., Penna, J., and Rombach, R.}
\newblock {SDXL:} improving latent diffusion models for high-resolution image synthesis.
\newblock In {\em The Twelfth International Conference on Learning Representations, {ICLR} 2024, Vienna, Austria, May 7-11, 2024\/} (2024).

\bibitem{Radford2021OpenAIClip}
{\sc Radford, A., Kim, J.~W., Hallacy, C., Ramesh, A., Goh, G., Agarwal, S., Sastry, G., Askell, A., Mishkin, P., Clark, J., Krueger, G., and Sutskever, I.}
\newblock Learning transferable visual models from natural language supervision.
\newblock In {\em ICML\/} (2021).

\bibitem{Rombach_2022_stable}
{\sc Rombach, R., Blattmann, A., Lorenz, D., Esser, P., and Ommer, B.}
\newblock High-resolution image synthesis with latent diffusion models.
\newblock In {\em Proceedings of the IEEE/CVF Conference on Computer Vision and Pattern Recognition (CVPR)\/} (June 2022), pp.~10684--10695.

\bibitem{U-Net}
{\sc Ronneberger, O., Fischer, P., and Brox, T.}
\newblock U-net: Convolutional networks for biomedical image segmentation.
\newblock In {\em Medical Image Computing and Computer-Assisted Intervention -- MICCAI 2015\/} (Cham, 2015), N.~Navab, J.~Hornegger, W.~M. Wells, and A.~F. Frangi, Eds., Springer International Publishing, pp.~234--241.

\bibitem{ruiz2022dreambooth}
{\sc Ruiz, N., Li, Y., Jampani, V., Pritch, Y., Rubinstein, M., and Aberman, K.}
\newblock Dreambooth: Fine tuning text-to-image diffusion models for subject-driven generation.
\newblock In {\em 2023 IEEE/CVF Conference on Computer Vision and Pattern Recognition (CVPR)\/} (Los Alamitos, CA, USA, 6 2023), IEEE Computer Society, pp.~22500--22510.

\bibitem{ILSVRC15}
{\sc Russakovsky, O., Deng, J., Su, H., Krause, J., Satheesh, S., Ma, S., Huang, Z., Karpathy, A., Khosla, A., Bernstein, M., Berg, A.~C., and Fei-Fei, L.}
\newblock {ImageNet Large Scale Visual Recognition Challenge}.
\newblock {\em International Journal of Computer Vision (IJCV) 115}, 3 (2015), 211--252.

\bibitem{schuhmann2022laion5bopenlargescaledataset}
{\sc Schuhmann, C., Beaumont, R., Vencu, R., Gordon, C.~W., Wightman, R., Cherti, M., Coombes, T., Katta, A., Mullis, C., Wortsman, M., Schramowski, P., Kundurthy, S.~R., Crowson, K., Schmidt, L., Kaczmarczyk, R., and Jitsev, J.}
\newblock {LAION}-5b: An open large-scale dataset for training next generation image-text models.
\newblock In {\em Thirty-sixth Conference on Neural Information Processing Systems Datasets and Benchmarks Track\/} (2022).

\bibitem{shuttleai}
{\sc ShuttleAI}.
\newblock {shuttleai/shuttle-3-diffusion}.
\newblock \url{https://huggingface.co/shuttleai/shuttle-3-diffusion/blob/main/vae/config.json}, 2024.
\newblock Accessed: May 27, 2025.

\bibitem{sohl2015deep}
{\sc Sohl-Dickstein, J., Weiss, E., Maheswaranathan, N., and Ganguli, S.}
\newblock Deep unsupervised learning using nonequilibrium thermodynamics.
\newblock In {\em International conference on machine learning\/} (2015), PMLR, pp.~2256--2265.

\bibitem{DDIM}
{\sc Song, J., Meng, C., and Ermon, S.}
\newblock Denoising diffusion implicit models.
\newblock In {\em International Conference on Learning Representations\/} (2021).

\bibitem{dreamstudio}
{\sc {Stability AI}}.
\newblock {DreamStudio - Image generation using Stable Diffusion}.
\newblock \url{https://beta.dreamstudio.ai/generate}.
\newblock Accessed: May 27, 2025.

\bibitem{Szegedy_2016_Inception}
{\sc Szegedy, C., Vanhoucke, V., Ioffe, S., Shlens, J., and Wojna, Z.}
\newblock Rethinking the inception architecture for computer vision.
\newblock In {\em Proceedings of the IEEE Conference on Computer Vision and Pattern Recognition (CVPR)\/} (6 2016).

\bibitem{tancik2020stegastamp}
{\sc Tancik, M., Mildenhall, B., and Ng, R.}
\newblock Stegastamp: Invisible hyperlinks in physical photographs.
\newblock In {\em Proceedings of the IEEE/CVF Conference on Computer Vision and Pattern Recognition (CVPR)\/} (6 2020).

\bibitem{tirkel1993electronic}
{\sc Tirkel, A.~Z., Rankin, G., Van~Schyndel, R., Ho, W., Mee, N., and Osborne, C.~F.}
\newblock Electronic watermark.
\newblock {\em Digital Image Computing, Technology and Applications (DICTA’93)\/} (1993), 666--673.

\bibitem{wen2023treering}
{\sc Wen, Y., Kirchenbauer, J., Geiping, J., and Goldstein, T.}
\newblock Tree-rings watermarks: Invisible fingerprints for diffusion images.
\newblock In {\em 37th Conference on Neural Information Processing Systems (NeurIPS)\/} (2023).

\bibitem{yang2024gaussian}
{\sc Yang, Z., Zeng, K., Chen, K., Fang, H., Zhang, W., and Yu, N.}
\newblock Gaussian shading: Provable performance-lossless image watermarking for diffusion models.
\newblock In {\em Proceedings of the IEEE/CVF Conference on Computer Vision and Pattern Recognition\/} (2024), pp.~12162--12171.

\bibitem{yu2021artificial}
{\sc Yu, N., Skripniuk, V., Abdelnabi, S., and Fritz, M.}
\newblock Artificial fingerprinting for generative models: Rooting deepfake attribution in training data.
\newblock In {\em Proceedings of the IEEE/CVF International conference on computer vision\/} (2021), pp.~14448--14457.

\bibitem{zeng2023securing}
{\sc Zeng, Y., Zhou, M., Xue, Y., and Patel, V.~M.}
\newblock Securing deep generative models with universal adversarial signature.
\newblock {\em arXiv preprint arXiv:2305.16310\/} (2023).

\bibitem{zhang2019robust}
{\sc Zhang, K.~A., Xu, L., Cuesta-Infante, A., and Veeramachaneni, K.}
\newblock Robust invisible video watermarking with attention.
\newblock {\em arXiv preprint arXiv:1909.01285\/} (2019).

\bibitem{zhang2018perceptual}
{\sc Zhang, R., Isola, P., Efros, A.~A., Shechtman, E., and Wang, O.}
\newblock The unreasonable effectiveness of deep features as a perceptual metric.
\newblock In {\em CVPR\/} (2018).

\bibitem{zhao2024sok}
{\sc Zhao, X., Gunn, S., Christ, M., Fairoze, J., Fabrega, A., Carlini, N., Garg, S., Hong, S., Nasr, M., Tramer, F., et~al.}
\newblock Sok: Watermarking for ai-generated content.
\newblock {\em arXiv preprint arXiv:2411.18479\/} (2024).

\bibitem{zhao2023invisible}
{\sc Zhao, X., Zhang, K., Su, Z., Vasan, S., Grishchenko, I., Kruegel, C., Vigna, G., Wang, Y.-X., and Li, L.}
\newblock Invisible image watermarks are provably removable using generative ai, 2023.

\bibitem{zhu2018hidden}
{\sc Zhu, J., Kaplan, R., Johnson, J., and Fei-Fei, L.}
\newblock Hidden: Hiding data with deep networks.
\newblock In {\em Proceedings of the European conference on computer vision (ECCV)\/} (2018), pp.~657--672.

\end{thebibliography}
